\documentstyle[12pt,epsf]{article}

\makeatletter
  
  \@addtoreset{equation}{section}
\makeatother

\begin{document}

%
%
\begin{titlepage}

\begin{flushright}
OU-HET 317 \\
hep-th/9903259 \\
March 1999
\end{flushright}

\bigskip

\begin{center}
{\Large \bf THREE-DIMENSIONAL BLACK HOLES} \\
{\Large \bf AND } \\
{\Large \bf LIOUVILLE FIELD THEORY}

\bigskip
\bigskip
T. Nakatsu, 
H. Umetsu 
and 
N. Yokoi \\
\bigskip
{\small \it
Department of Physics,\\
Graduate School of Science, Osaka University,\\
Toyonaka, Osaka 560, JAPAN
}
\end{center}

\bigskip
\bigskip

\begin{abstract}
A quantization of $(2+1)$-dimensional gravity 
with negative cosmological constant is presented 
and quantum aspects of the $(2+1)$-dimensional 
black holes are studied thereby. 
The quantization consists of two procedures. 
One is related with quantization of the asymptotic 
Virasoro symmetry.  
A notion of the Virasoro deformation of $3$-geometry 
is introduced. For a given black hole,  
the deformation of the exterior of the outer horizon 
is identified with a product of appropriate coadjoint orbits 
of the Virasoro groups $\widehat{{\it diff}S^1}_{\pm}$. 
Its quantization provides 
unitary irreducible representations of the Virasoro algebra, 
in which state of the black hole becomes primary.   
To make the quantization complete, 
${\it holonomies}$, the global degrees of freedom, 
are taken into account.  
By an identification of these topological operators 
with zero modes of the Liouville field,   
the aforementioned unitary representations reveal, 
as far as $c \gg 1$, 
as the Hilbert space of this two-dimensional 
conformal field theory. 
This conformal field theory, 
living on the cylinder at infinity of the black hole 
and having continuous spectrums,  
can recognize the outer horizon only as 
a {\it one-dimensional object} in $SL_2({\bf R})$ 
and realize it as insertions of the corresponding 
vertex operator. Therefore it can not be 
a conformal field theory on the horizon. 
Two possible descriptions of the horizon conformal 
field theory are proposed. 

\end{abstract}

\end{titlepage}



%
%

\section{Introduction}

       It has been proposed 
\cite{Maldacena,G-K-P,Witten3} 
that supergravity (and string theory) on 
$AdS_{d+1}$ times a compact manifold 
is equivalent to a conformal field theory 
living on the boundary of $AdS_{d+1}$. 
Supergravity (and string theory) on 
asymptotically $AdS_{d+1}$ space-time is also expected 
\cite{Witten3}
to be realized by a boundary field theory on a suitable 
vacuum.

     Investigations of the bulk quantum gravity,   
in particular, to clarify $AdS$ black holes 
as the quantum gravitational states,        
will lead a deeper insight on our understanding 
of the $AdS/CFT$ correspondence.  In three dimensions,  
a description of gravity in terms of gauge theory 
\cite{Witten1} is known, 
by which quantum gravity in three dimensions seems 
to be tractable and thus consideration about such quantum 
gravitational states will become possible.

      In this article we present a quantization 
of $(2+1)$-dimensional gravity with negative cosmological 
constant $\Lambda=-1/l^2$ and thereby 
investigate quantum aspects 
of the $(2+1)$-dimensional black holes 
from the perspective of the $AdS_3/CFT_2$ correspondence.

       The quantization consists 
of  the following two procedures. 
The first is related with quantization of the 
asymptotic Virasoro symmetry found by Brown and Henneaux 
\cite{Brown-Henneaux}. 
To describe it we introduce 
in Section 2 a notion of the Virasoro 
deformation of three-geometry. 
It is originally defined as a deformation of a suitable 
non-compact simply-connected region in $SL_2({\bf R})$ 
and is given by the coadjoint action 
of the Virasoro groups 
$\widehat{{\it diff}S^1}_+ \times \widehat{{\it diff}S^1}_-$.  
Induced metric from $SL_2({\bf R})$ 
provides a metric of the deformed region. 
Given a black hole, we can take,  
as such a non-compact simply-connected region,  
a covering space of the exterior of the outer horizon. 
Its Virasoro deformation commutes with the projection. 
Therefore we can obtain a family of the deformed quotients. 
This defines the Virasoro deformation 
of the exterior of the outer horizon of the black hole. 
The induced metric of the deformed region 
becomes that of its quotient, which can be 
understood as a deformed metric of the black hole 
obtained in \cite{Banados1}. 
The asymptotic Virasoro symmetry turns out 
to be the infinitesimal form of this deformation. 
The family of the deformed quotients 
will be identified with a product of 
the coadjoint orbits of 
the Virasoro groups $\widehat{{\it diff}S^1}_{\pm}$ 
labeled by mass and angular momentum of the black hole 
together with the cosmological constant. 
In Section 3 we provide quantization of 
the asymptotic Virasoro symmetry or 
the Virasoro deformation of the exterior of the outer horizon.   
It is prescribed 
by quantization of the corresponding coadjoint orbits. 
Quantization of each orbit turns out to give  
a unitary irreducible representation of 
the Virasoro algebra with central charge 
$c=\frac{3l}{2G}$.  
State of the black hole becomes   
the primary state of the representation.  
All the excited states of the representation 
correspond to the local excitations 
by the Virasoro deformation, that is, 
as noted in \cite{Martinec}, 
gravitons.   
$AdS_3$ is also examined. Quantization of its Virasoro 
deformation turns out to provide 
a unitary irreducible representation, 
in which state of $AdS_3$ is  
the $SL_2({\bf R})_+\times SL_2({\bf R})_-$-invariant 
primary state.

                Although the quantization of the Virasoro 
deformations leads the unitary irreducible representations 
of the Virasoro algebra, it is not sufficient as a  
quantization of three-dimensional gravity. 
If one takes the perspective of 
$SL_2({\bf R})_+\times SL_2({\bf R})_-$ 
Chern-Simons gravity \cite{Witten1}, 
the Virasoro deformation corresponds to 
local degrees of freedom of the theory 
and, 
in order to make the quantization complete, 
we must take into account of the global degrees of freedom, 
{\it i.e.}, holonomies. 
We start Section 4 by discussing their quantization.  
Having quantized holonomies, 
it is very reasonable to expect from the perspective of the
$AdS_{3}/CFT_{2}$ correspondence that, by an 
identification of these topological operators 
with the zero modes of a suitable two-dimensional 
quantum field, the unitary irreducible representations 
obtained by the quantization of the Virasoro deformations 
could be reproduced as the Hilbert space of the two-dimensional 
conformal field theory. This expectation turns out to 
be true at least when $c \gg 1$. The holonomy variables 
will be identified with the zero modes of a real scalar 
field $X$, which is interpreted as the Liouville field. 
The state of $AdS_3$ is identified 
with the $sl_2({\bf C})$-invariant 
vacuum of this conformal field theory. 
The states of the black holes are the primary states 
obtained from the vacuum by an operation 
of the corresponding vertex operators.  
Excitations of the Liouville field are equivalent with 
the generators of the Virasoro deformation. 
This field theory admits to have continuous spectrums as 
expected.

           Having obtained the quantization of three-dimensional 
gravity by the Liouville field theory, 
the following question may arise : 
How can one understand the three-dimensional black holes  
in two dimensions where the Liouville field lives ? 
The Liouville field theory is living on a two-sphere 
which is obtained by a compactification of the boundary 
cylinder at infinity of the black holes. 
To answer this question it becomes necessary to 
understand  how the Virasoro deformation recognizes  
the outer horizon of the black hole. 
It will be shown at the end of Section 4 that 
it recognizes the outer horizon not as a two-dimensional 
object but as a one-dimensional object in $SL_2({\bf R})$. 
Therefore the boundary Liouville field $X$ 
can see the outer horizon only as a one-dimensional object. 
Compactification of the boundary cylinder to a sphere 
simultaneously makes the solid cylinder  
to a three-ball, by which this one-dimensional outer horizon 
intersects with the boundary sphere at two points. 
The boundary Liouville field theory recognizes 
its intersections as insertions of 
the corresponding vertex operator.

         A microscopic description of the three-dimensional 
black holes has been proposed by 
Carlip \cite{Carlip1} and 
Strominger \cite{Strominger1,Maldacena-Strominger} using 
a conformal field theory on the horizon. 
Since the Virasoro deformation of the exterior of the outer 
horizon can not recognize the horizon as a two-dimensional 
object, the boundary Liouville field theory can not be equivalent to 
the horizon conformal field theory. 
In Section 5 we propose two possible descriptions of the 
horizon conformal field theory. 
One is a string theory in the background of a macroscopic string 
living in $SL_2({\bf R})$.  
Worldvolume of the macroscopic string should be  
identified with the two-dimensional outer horizon, 
by which we can obtain the Virasoro algebra on the horizon 
as its fluctuation by microscopic string.  
The other is a conformal field theory which could be 
obtained by the Virasoro deformation of the region 
between the inner and outer horizons 
of the black hole. The second possibility seems to imply an  
interpretation of the conjecture 
made by Maldacena \cite{Maldacena},  
in terms of a gauge transformation.

\section{More On Geometry In Three Dimensions}

The BTZ black holes
\footnote{Exact solutions of the vacuum 
Einstein equation with a negative cosmological 
constant $\Lambda=-1/l^2$.} \cite{BTZ1} 
are three-dimensional black holes specified 
by their mass and angular momenta.  
The BTZ black hole with mass $M (\geq 0)$ and angular momentum $J$ 
will be denoted by $X_{(J,M)}$. One can avoid a naked singularity 
by imposing the bound, $|J| \leq Ml$, on the allowed value of 
its angular momentum.   
In terms of the Schwarzschild coordinates 
$(t,\phi,r)$, where the ranges are taken 
$-\infty<t<+\infty$, 
$0\leq \phi <2\pi$ 
and $0<r<+\infty$, 
the black hole metric $ds_{X_{(J,M)}}^2$ 
has the form 
\begin{eqnarray}
ds^{2}_{X_{(J,M)}} 
\equiv 
-N^2(dt)^2+N^{-2}(dr)^2
+r^2(d\phi+N^{\phi}dt)^2 . 
\label{dsX(J,M)}
\end{eqnarray}
$N$ and $N^{\phi}$ are functions of the radial coordinate $r$ 
and given by 
\begin{eqnarray}
N^2 =
\frac{(r^2-r_+^2)(r^2-r_-^2)}{l^2r^2},~~~~
N^{\phi} =  
\left\{ 
\begin{array}{cc}
\frac{r_+r_-}{lr^2} & \mbox{when}~~J \geq 0, 
\\ 
-\frac{r_+r_-}{lr^2} & \mbox{when}~~J < 0.  
\end{array}
\right. 
\label{N-Nphi}
\end{eqnarray}
The outer and inner horizons are located respectively 
at $r=r_+$ and $r=r_-$. Information about the mass 
and angular momentum is encoded in $r_{\pm}$ by  
\begin{eqnarray} 
r_{\pm}^2 \equiv 
4GMl^2 \left( 
1\pm \sqrt{1-\frac{J^2}{M^2l^2}} 
\right), 
\label{rpm}
\end{eqnarray}
where $G$ is Newton constant. 
In the case of $Ml=|J|$,  the black hole is called extremal. 
It holds $r_+=r_-$  
and then the outer and inner horizons coincide  
with each other. 
When $Ml>|J|$, it is called non-extremal.

\subsection{More on the black hole}

The exterior of the outer horizon 
$r=r_+$ of the BTZ black hole will play an important role 
throughout this paper. 
We denote 
the exterior of the outer horizon of 
$X_{(J,M)}$ by $X_{(J,M)}^+$. 
That is,  
$X_{(J,M)}^+\equiv 
\left \{  \left. 
(t,\phi,r)\in X_{(J,M)} 
\right | r>r_+  \right \}$. 
It is known \cite{BTZ1,BTZ2} that the black hole can 
be identified with an appropriate quotient 
of $SL_2({\bf R})$.
We will need a further detailed description of this relation 
between $X_{(J,M)}^+$ and $SL_2({\bf R})$. 
In this subsection we provide it  
for the case of the non-extremal black holes, 
that is, the case of $Ml > |J|$. 
Discussion presented here becomes a prototype for 
the other cases, which will be treated briefly in the 
next section.

            $SL_2({\bf R})$ is the three-dimensional 
non-compact group manifold admitted to have invariant 
metrics which scalar curvatures are negative constant. 
These metrics are the Killing metric and its rescalings. 
Among them, let $ds_{SL_2({\bf R})}^2$ be the metric 
which scalar curvature equals to $-6/l^2$. It is given by 
\begin{eqnarray}
ds^2_{SL_2({\bf R})}\equiv \frac{l^2}{2}
{\rm Tr}~ (g^{-1}dg)^2 . 
\label{ds SL2}
\end{eqnarray}   
This metric is an exact solution of the vacuum Einstein 
equation with the negative cosmological constant 
$\Lambda=-1/l^2$.  
Any element $g$ of $SL_2({\bf R})$ can be written as 
$g=e^{-(\tau+\theta)J^0}e^{\sigma J^1}e^{-(\tau-\theta)J^0}$, 
where $0 \leq \tau,\theta <2\pi$ and $0\leq \sigma <+\infty$
\footnote{For $SL_2({\bf R})$, a basis of the Lie algebra over 
${\bf R}$ is taken by the three matrices $J^a$ 
\begin{eqnarray}
J^0 \equiv
\left( \begin{array}{cc}
0 & \frac{1}{2} \\
-\frac{1}{2} & 0 
\end{array} \right),~~
J^1 \equiv 
\left( \begin{array}{cc}
-\frac{1}{2} & 0 \\
0 &  \frac{1}{2}  
\end{array} \right),~~
J^2 \equiv 
\left( \begin{array}{cc}
0 & \frac{1}{2} \\
\frac{1}{2} & 0 
\end{array} \right).
\nonumber 
\end{eqnarray} 
These enjoy 
$\left[ J^a,J^b \right]=
\epsilon^{abc}J_c$, 
where $\epsilon^{abc}$ is 
a totally anti-symmetric tensor 
normalized by $\epsilon^{012}=1$ and   
the $sl_2({\bf R})$-indices are 
lowered (or raised) by $\eta_{ab}\equiv diag.(-1,1,1)$ 
(or $\eta^{ab}$, the inverse matrix of $\eta_{ab}$).}. 
With this Cartan decomposition one can see that 
$SL_2({\bf R})$ is topologically a solid torus. 
See Fig.1.
\begin{figure}[t]
\epsfysize=5cm
\makebox[17cm][c]{\epsfbox{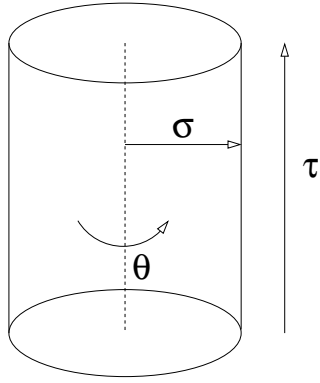}}
\caption{{\small $SL_2({\bf R})$ is a solid torus. The upper and lower sides
are identified.}}
\end{figure}

          Let $Z_0$ be a non-compact 
simply-connected region of $SL_2({\bf R})$ 
consisting of the group elements
\begin{eqnarray}
e^{-\varphi J^2}
e^{\sigma J^1}
e^{\psi J^2}, 
\label{Z0 non-ex}
\end{eqnarray}
where $\varphi, \psi$ and $\sigma$ are real parameters. 
Their ranges are taken $-\infty < \varphi,\psi < +\infty$ and 
$0 < \sigma < +\infty$. 
$Z_{0}$ is illustrated in Fig.2.
\begin{figure}[t]
\epsfysize=8cm
\makebox[17cm][c]{\epsfbox{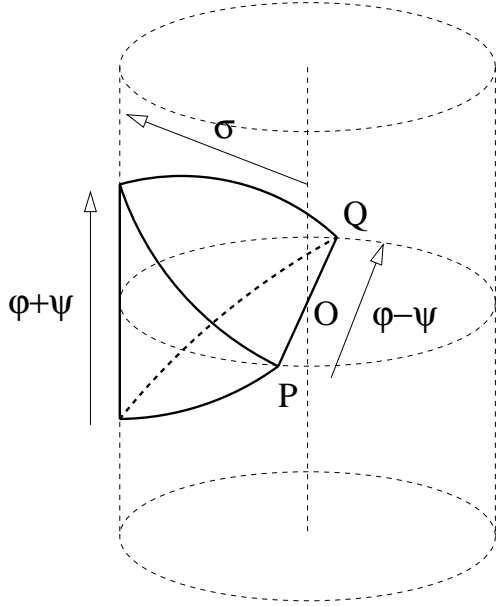}}
\caption{{\small The non-compact simply-connected region 
$Z_{0}$ of $SL_2({\bf R})$}}
\end{figure}
Notice that the line $POQ$ ($\sigma=0$) in Fig.2 
is excluded by the definition.   
The invariant metric (\ref{ds SL2}) 
defines an induced metric on $Z_0$. 
We call it  $ds_0^2$. 
It is merely a restriction of (\ref{ds SL2}) on $Z_0$. 
Regarding the parameters $\varphi,\psi$ and $\sigma$ 
as coordinates of $Z_0$, 
it can be written in the form 
\begin{eqnarray}
ds^2_0=
\frac{l^2}{4}
\left \{(d\varphi)^2+(d\psi)^2 
-(e^{\sigma}+e^{-\sigma})d\varphi d\psi 
+(d\sigma)^2 \right \}.
\label{ds0 non-ex}
\end{eqnarray}

                         We can provide a description 
of the above region together with the induced metric 
in terms of $SL_2({\bf R})_+\times SL_2({\bf R})_-$ 
Chern-Simons gravity. 
To present the description, 
it is convenient to introduce a mapping,~ 
$SL_2({\bf R})_+\times SL_2({\bf R})_- 
\rightarrow SL_2({\bf R})$,~ by 
\begin{eqnarray} 
(g^{(+)},g^{(-)})~ \longmapsto~ 
{g^{(+)}}^{-1}\! g^{(-)}. 
\label{pairing of SL2pm}
\end{eqnarray}
This defines a pairing of $SL_2({\bf R})_{\pm}$ 
which is invariant under the left diagonal 
$SL_2({\bf R})$ action,~ 
$g^{(\pm)}\mapsto gg^{(\pm)}$. 
It is also convenient to regard 
the group elements (\ref{Z0 non-ex}) 
as an image of a function $h$ which takes its values 
in $SL_2({\bf R})$ as presented in (\ref{Z0 non-ex}). 
Explicitly $h$ is given by  
\begin{eqnarray}
h(\varphi,\psi,\sigma) \equiv 
e^{-\varphi J^2}
e^{\sigma J^1}
e^{\psi J^2}. 
\label{h non-ex}
\end{eqnarray}
With this perspective, 
we can consider the following decomposition of $h$ 
into a product of $SL_2({\bf R})_{\pm}$-valued functions 
$h^{(\pm)}$ : 
\begin{eqnarray}
h={h^{(+)}}^{-1}h^{(-)}. 
\label{def of hpm}
\end{eqnarray}
This decomposition is equivalent 
to construct a pre-image of $Z_0$ 
in the mapping (\ref{pairing of SL2pm}). 
It is unique up to the left diagonal $SL_2({\bf R})$ 
gauge transformation. 
This symmetry turns out to be the three-dimensional 
local Lorentz symmetry.
Given such $SL_2({\bf R})_{\pm}$-valued functions, 
we can associate flat $SL_2({\bf R})$ connections 
$A_0^{(\pm)}$ as follows :  
\begin{eqnarray}
A_0^{(\pm)}\equiv h^{(\pm)}d{h^{(\pm)}}^{-1}. 
\label{def of A0}
\end{eqnarray} 
It turns out that 
$A_0 \equiv (A_0^{(+)},A_0^{(-)})$ is a desired 
classical solution of the Chern-Simons gravity. 
In fact, 
one can reproduce the metric (\ref{ds0 non-ex}) 
by means of the 3-vein $e^a$ which is obtained from 
$A_0$ through the following correspondence 
\cite{AT,Witten1} 
between the relativity and gauge theory languages:
\begin{eqnarray}
A_0^{(\pm)}=\pm \frac{1}{l}e+\omega.
\label{A0 by (e,w)}
\end{eqnarray} 
Both 3-vein $e^a$ and spin connection $\omega^a$ 
are realized as $sl_2({\bf R})-$valued one-forms,~  
$e \equiv e^a J_a$ and $\omega \equiv \omega^a J_a$.  
To be explicit, by use of this correspondence, 
the 3-vein $e$ has the expression  
$e=\frac{l}{2}(A_0^{(+)}-A_0^{(-)})$.  
It can be also expressed as   
$e=\frac{l}{2}h^{(-)}\cdot {h}^{-1}
dh \cdot {h^{(-)}}^{-1}$. 
Then the metric $e^ae_a$ ~$(=2~{\rm Tr}~e^2)$~  
acquires the form,  
$\frac{l^2}{2} {\rm Tr}~ (h^{-1}dh)^2$.  
This is exactly the induced metric $ds_0^2$. 
The correspondence (\ref{A0 by (e,w)}) also shows 
that the flatness of the connection $A_0$  
is equivalent to the Cartan form of the vacuum Einstein equation 
with the negative cosmological constant  
\begin{eqnarray}
de^a+\epsilon^{abc}\omega_b \wedge e_c &=& 0, 
\nonumber \\ 
dw^a+\frac{1}{2}\epsilon^{abc}\omega_b \wedge \omega_c 
+\frac{1}{2l^2}\epsilon^{abc}e_b \wedge e_c &=& 0. 
\end{eqnarray}

         The non-compact simply-connected region $Z_0$ 
is a covering space of $X_{(J,M)}^+$,   
the exterior of the outer horizon 
of the non-extremal black hole $X_{(J,M)}$.      
The projection $\pi$ can be described 
by using the relations  
\begin{eqnarray}
\frac{r_++r_-}{l}\left(\frac{t}{l}+\phi \right)
&=&\varphi,
\nonumber \\ 
\frac{r_+-r_-}{l}\left(\frac{t}{l}-\phi \right)
&=&\psi, 
\nonumber \\ 
\frac{\sqrt{r^2-r_+^2}+\sqrt{r^2-r_-^2}}
{\sqrt{r_+^2-r_-^2}}
&=&e^{\sigma/2},
\label{pi non-ex}
\end{eqnarray}
besides the identification of the Schwarzschild angular coordinate,  
$\phi \sim \phi+2\pi$. 
See Fig.3.
\begin{figure}[t]
\epsfysize=8cm
\makebox[17cm][c]{\epsfbox{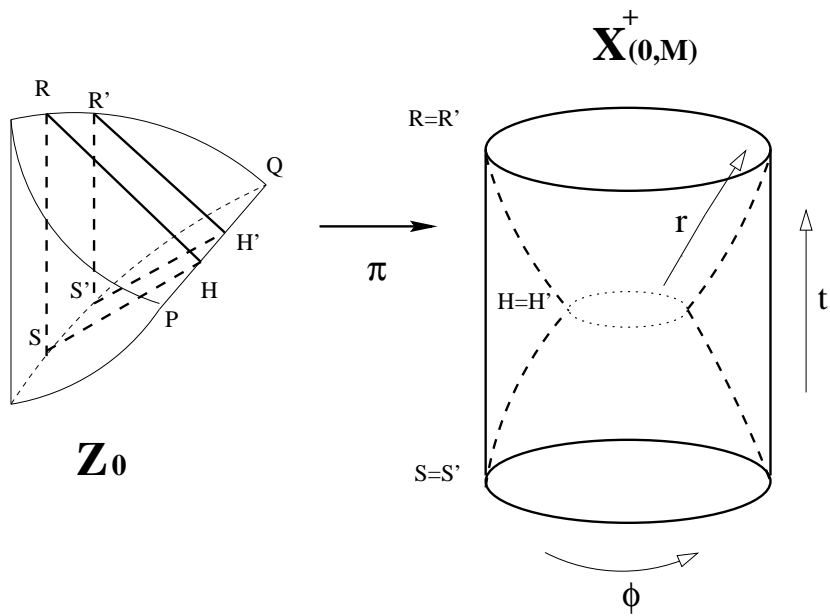}}
\caption{{\small The projection from $Z_{0}$ to the exterior of the outer
horizon of the non-extremal black hole. $J = 0$ case
is depicted as an example. The space between two surfaces $HRS$ and
$H^{'}R^{'}S^{'}$ is the fundamental region.}}
\end{figure}
This projection map becomes the isometry : 
$\pi_* ds_{0}^2=ds^2_{X_{(J,M)}^+}$. 
Therefore the exterior of the outer horizon can be identified with 
the quotient of $Z_0$ :  
\begin{eqnarray}
X^+_{(J,M)}=Z_0/\sim. 
\label{quotient non-ex}
\end{eqnarray}
Here the equivalence relation ``$\sim$", 
which originates in the periodicity of $\phi$, is given by  
\begin{eqnarray}
(\varphi,\psi,\sigma)\sim 
(\varphi+2\pi\frac{r_++r_-}{l},\psi-2\pi\frac{r_+-r_-}{l},\sigma). 
\label{equiv non-ex}
\end{eqnarray}

      It turns out convenient to introduce another parametrization 
of $Z_0$ which is slightly different from $(\varphi,\psi,\sigma)$. 
Let $b_0,\tilde{b}_0$ and $c$ be the quantities 
\begin{eqnarray}
b_0\equiv \frac{(r_++r_-)^2}{16Gl},~~~
\tilde{b}_0 \equiv \frac{(r_+-r_-)^2}{16Gl},~~~
c \equiv \frac{3l}{2G}. 
\label{b0-tb0-c}
\end{eqnarray}  
We can write any element of $Z_0$ in the form  
\begin{eqnarray}
e^{-\sqrt{\frac{b_0}{c/24}}wJ^2}
e^{(\rho-\frac{1}{2}\ln \frac{b_0\tilde{b}_0}{(c/6)^2})J^1}
e^{\sqrt{\frac{\tilde{b}_0}{c/24}}\tilde{w}J^2},  
\label{Z0 non-ex 2}
\end{eqnarray} 
where $w,\tilde{w}$ and $\rho$ are real parameters. 
To exactly cover $Z_0$ with this parametrization, 
the ranges of $w,\tilde{w}$ and $\rho$ 
must be taken as $-\infty < w,\tilde{w} <+\infty$ and 
$\frac{1}{2} \ln \frac{b_0\tilde{b}_0}{(c/6)^2}
<\rho<+\infty$.

               Using this parametrization or coordinates 
of $Z_0$, let us provide an explicit form of the corresponding 
classical solution of the Chern-Simons gravity.  
The $SL_2({\bf R})$-valued function $h$, which is 
defined in (\ref{h non-ex}), 
can be read in this coordinate as 
\begin{eqnarray}
h(w,\tilde{w},\rho)=
e^{-\sqrt{\frac{b_0}{c/24}}wJ^2}
e^{(\rho-\frac{1}{2}\ln \frac{b_0\tilde{b}_0}{(c/6)^2})J^1}
e^{\sqrt{\frac{\tilde{b}_0}{c/24}}\tilde{w}J^2}. 
\label{h non-ex 2}
\end{eqnarray}
The following decomposition of (\ref{h non-ex 2}) 
may be chosen  
\begin{eqnarray}
h^{(+)}&=&
e^{-\frac{1}{2}
(\rho-\ln \frac{b_0}{c/6})J^1}
e^{\sqrt{\frac{b_0}{c/24}}wJ^2},
\nonumber \\ 
h^{(-)}&=&
e^{\frac{1}{2}
(\rho-\ln 
\frac{\tilde{b}_0}{c/6})J^1}
e^{\sqrt{\frac{\tilde{b}_0}{c/24}}\tilde{w}J^2}.
\label{hpm non-ex}
\end{eqnarray}
With this choice, 
the flat connections $A_0^{(\pm)}$, 
which are defined in (\ref{def of A0}), become  
\begin{eqnarray}
A_0^{(+)}=
\left( \begin{array}{cc}
-\frac{1}{4}d\rho & -e^{\rho/2}dw \\
-\frac{b_0}{c/6}e^{-\rho/2}dw & \frac{1}{4}d\rho  
\end{array} \right),~~~
A_0^{(-)}=
\left( \begin{array}{cc}
\frac{1}{4}d\rho & 
-\frac{\tilde{b}_0}{c/6}e^{-\rho/2}d\tilde{w} \\
-e^{\rho/2}d\tilde{w} & -\frac{1}{4}d\rho  
\end{array} \right).
\label{A0 non-ex}
\end{eqnarray}
Then $A_0=(A_0^{(+)},A_0^{(-)})$ turns out to 
give the metric  
\begin{eqnarray}
ds_{(b_0,\tilde{b_0})}^2=
l^2
\left \{
\frac{b_0}{c/6}(dw)^2+
\frac{\tilde{b}_0}{c/6}(d\tilde{w})^2
-\left(e^{\rho}+
\frac{b_0\tilde{b}_0}{(c/6)^2}e^{-\rho}
\right)
dwd\tilde{w}
+\frac{1}{4}(d\rho)^2\right \}.
\label{ds0-2}
\end{eqnarray} 
After the change of the coordinates, 
which can be read from 
(\ref{Z0 non-ex}) and (\ref{Z0 non-ex 2}),  
it coincides with the induced metric $ds^2_0$.  
In terms of the coordinates $(w,\tilde{w},\rho)$, 
the equivalence relation (\ref{equiv non-ex}) used in 
the identification with the black hole becomes simple :
\begin{eqnarray}
(w,\tilde{w},\rho)\sim 
(w+2\pi,\tilde{w}-2\pi,\rho).
\label{equiv non-ex2}
\end{eqnarray}
Therefore we can also say that  
the quotient of $Z_0$ 
by the identification (\ref{equiv non-ex2}), 
together with the induced metric (\ref{ds0-2}),   
is identified with the exterior of the outer horizon of $X_{(J,M)}$.


\subsection{Generalization of flat connection}

   A generalization of the flat 
$SL_2({\bf R})_+\times SL_2({\bf R})_-$ connection (\ref{A0 non-ex}) 
is examined \cite{CHD-BO,Banados1} to find a realization of 
the asymptotic Virasoro symmetry \cite{Brown-Henneaux} 
in terms of the Chern-Simons gravity. 
It has the following form 
\begin{eqnarray}
A_b^{(+)}=
\left( \begin{array}{cc}
-\frac{1}{4}d\rho & -e^{\rho/2}dw \\
-\frac{b(w)}{c/6}e^{-\rho/2}dw & \frac{1}{4}d\rho  
\end{array} \right),~~~
A_{\tilde{b}}^{(-)}=
\left( \begin{array}{cc}
\frac{1}{4}d\rho & 
-\frac{\tilde{b}(\tilde{w})}{c/6}e^{-\rho/2}d\tilde{w} \\
-e^{\rho/2}d\tilde{w} & -\frac{1}{4}d\rho  
\end{array} \right).
\label{A(b,tb)}
\end{eqnarray} 
The suffix $b$ and $\tilde{b}$ label the flat connections.
To be consistent with the equivalence relation (\ref{equiv non-ex2}), 
we assume that $b$ and $\tilde{b}$ are respectively real functions of 
$w$ and $\tilde{w}$ with $2\pi$ periodicity.

          The asymptotic Virasoro algebra may be realized 
by the following generators of the infinitesimal gauge transformation 
\begin{eqnarray}
\lambda_f^{(+)}=
\left( \begin{array}{cc}
-\frac{1}{2}f' & fe^{\rho/2} \\
(\frac{b}{c/6}f-\frac{1}{2}f'')e^{-\rho/2} & \frac{1}{2}f' 
\end{array} \right),~~~
\lambda_{\tilde{f}}^{(-)}=
\left( \begin{array}{cc}
\frac{1}{2}\tilde{f}' & 
(\frac{\tilde{b}}{c/6}\tilde{f}-\frac{1}{2}\tilde{f}'')e^{-\rho/2} \\
\tilde{f}e^{\rho/2} & -\frac{1}{2}\tilde{f}'  
\end{array} \right), 
\end{eqnarray}
where, again to be consistent with (\ref{equiv non-ex2}), 
$f=f(w)$ and $\tilde{f}=\tilde{f}(\tilde{w})$ 
are assumed to be real functions with $2\pi$ periodicity.
The gauge transformation of the connection (\ref{A(b,tb)}) 
equals \cite{Polyakov} to deformations of $b$ and $\tilde{b}$
\begin{eqnarray}
\delta_{\lambda_f^{(+)}}A_b^{(+)}
=\left( \begin{array}{cc} 
0 & 0 \\ 
-\frac{\delta_fb}{c/6}e^{-\rho/2}dw & 0 
\end{array}\right),~~
\delta_{\lambda_{\tilde{f}}^{(-)}}A_{\tilde{b}}^{(-)}
=\left( \begin{array}{cc} 
0 & 
-\frac{\delta_{\tilde{f}}\tilde{b}}{c/6}
e^{-\rho/2}d\tilde{w} \\ 
0 & 0 
\end{array}\right). 
\label{Polyakov transform}
\end{eqnarray} 
These deformations have the forms 
\begin{eqnarray}
\delta_{f}b
&=&fb'+2f'b-\frac{c}{12}f''' ,
\nonumber \\ 
\delta_{\tilde{f}}\tilde{b} 
&=&
\tilde{f}\tilde{b}'+2\tilde{f}'\tilde{b}
-\frac{c}{12}\tilde{f}'''. 
\label{coadjoint action1} 
\end{eqnarray} 
In the asymptotic region where $\rho$ is very large, 
the non-zero off-diagonal elements of 
$\delta_{\lambda_{f}^{(+)}}A_{b}^{(+)}$ 
and 
$\delta_{\lambda_{\tilde{f}}^{(-)}}A_{\tilde{b}}^{(-)}$ 
become negligible. 
Hence the gauge transformation can 
be regarded as a asymptotic symmetry.  
To contact with the asymptotic Virasoro symmetry 
it is convenient to decompose the gauge transformation 
into the diffeomorphism and the local Lorentz transformation. 
Notice that, for a given flat connection $A$,  
its infinitesimal general coordinates transform  
(say, generated by a vector field $\xi$) 
can be interpreted \cite{Witten1} 
as its infinitesimal gauge transform. 
Via the Sugawara construction 
the generator of the gauge transformation is given by 
$\lambda \equiv \xi^{\mu}A_{\mu}$.
Decompose the gauge transformation 
(\ref{Polyakov transform}) to  
$\delta_{\lambda_{f}^{(+)}}A_{b}^{(+)}$
$=\delta_{\xi}A_{b}^{(+)}+\delta_{\eta}A_{b}^{(+)}$ and 
$\delta_{\lambda_{\tilde{f}}^{(-)}}A_{\tilde{b}}^{(-)}$ 
$=\delta_{\tilde{\xi}}A_{\tilde{b}}^{(-)}+\delta_{\eta}
A_{\tilde{b}}^{(-)}$,  
where $\xi,\tilde{\xi}$ are vector fields 
and $\eta$ is a generator of the local Lorentz transformation 
(the diagonal $SL_2({\bf R})$ gauge transformation). 
 The vector fields turn out to be 
\begin{eqnarray}
\xi
&=&2f'\partial_{\rho}
+\left(-f
+\frac{\frac{\tilde{b}}{c/6}
          f''e^{-\rho} +\tilde{f}''}
      {\frac{b\tilde{b}}{(c/6)^2}e^{-\rho} 
                     -e^{\rho}}\right)
      \partial_{w}, \nonumber \\ 
\tilde{\xi}&=&
2\tilde{f}'\partial_{\rho}
+\left(-\tilde{f}
+\frac{\frac{b}{c/6}\tilde{f}''e^{-\rho}+f''}
      {\frac{b\tilde{b}}{(c/6)^2}e^{-\rho} 
                     -e^{\rho}}\right)
      \partial_{\tilde{w}}.
\label{xi-barxi}
\end{eqnarray}
In the asymptotic region they behave as 
$\xi$ 
$\sim 2f'\partial_{\rho}-f\partial_{w}$ 
and 
$\tilde{\xi} \sim$ 
$2\tilde{f}'\partial_{\rho}-\tilde{f}\partial_{\tilde{w}}$. 
These coincide with the Virasoro generators 
considered in \cite{Brown-Henneaux}.


\subsection{Generalization of black hole}

    Given the flat connection (\ref{A(b,tb)}), 
one can construct a metric according to the prescription 
of the Chern-Simons gravity. Denote this metric by 
$ds_{(b,\tilde{b})}^2$. It has the form 
\begin{eqnarray}
ds^2_{(b,\tilde{b})}
=l^2
\left \{
\frac{b}{c/6}(dw)^2+
\frac{\tilde{b}}{c/6}(d\tilde{w})^2
-\left(e^{\rho}+
\frac{b\tilde{b}}{(c/6)^2}e^{-\rho}
\right)
dwd\tilde{w}
+\frac{1}{4}(d\rho)^2\right \}. 
\label{ds(b,tb)}
\end{eqnarray} 
It is also possible to obtain a region of $SL_2({\bf R})$,  
analogous to $Z_0$, 
on which the induced metric from $SL_2({\bf R})$ 
is precisely given by (\ref{ds(b,tb)}). 
Since $A_{b}^{(+)}$ and $A_{\tilde{b}}^{(-)}$ 
are flat connections we can always 
trivialize them by some $SL_2({\bf R})_{\pm}$-valued functions 
$h_{b}^{(+)}$ and $h_{\tilde{b}}^{(-)}$
\begin{eqnarray}
A_b^{(+)}=h_{b}^{(+)}d{h_{b}^{(+)}}^{-1},~~~
A_{\tilde{b}}^{(-)}=h_{\tilde{b}}^{(-)}d{h_{\tilde{b}}^{(-)}}^{-1}. 
\end{eqnarray}
Then the desired region of $SL_2({\bf R})$, 
which we will call $Z_{(b,\tilde{b})}$,     
consists of those obtained from 
$h_{b}^{(+)}$ and $h_{\tilde{b}}^{(-)}$ 
by their local Lorentz invariant pairing,  
${h_b^{(+)}}^{-1}h_{\tilde{b}}^{(-)}$.  
It is easy to see that the induced metric on 
$Z_{(b,\tilde{b})}$ is nothing but $ds_{(b,\tilde{b})}^2$.  
When both $b$ and $\tilde{b}$ are constant, 
$Z_{(b,\tilde{b})}$ coincides with $Z_0$.  
In particular $Z_{(b_0,\tilde{b}_0)}=Z_0$ holds. 
On the other hand, 
as far as $b$ or $\tilde{b}$ is not constant, 
$Z_{(b,\tilde{b})}$ becomes the region of $SL_2({\bf R})$ 
which is different from $Z_0$. One may suspect that  
a reparametrization of $Z_0$ such as 
$\displaystyle{
e^{-\int^w\sqrt{\frac{b}{c/24}}dwJ^2}  
e^{(\rho-\frac{1}{2}\ln \frac{b\tilde{b}}{(c/6)^2})J^1} 
e^{\int^{\tilde{w}}\sqrt{\frac{\tilde{b}}{c/24}}d\tilde{w}J^2}}$   
might lead the flat connection (\ref{A(b,tb)}) or the metric 
(\ref{ds(b,tb)}). 
But it does not so 
as far as $b$ or $\tilde{b}$ is not constant.

     It is an interesting but hard task to describe the 
region $Z_{(b,\tilde{b})}$ explicitly. Nevertheless, the following 
argument shows that $Z_{(b.\tilde{b})}$ is generically different from $Z_0$ 
in $SL_2({\bf R})$, which is enough for our discussion in this article. 
We first notice that the completion of $Z_0$ at $\sigma=0$ 
is achived by adding the one-dimensional line 
$POQ$ (Fig.2) in $SL_2({\bf R})$.
It is a counterpart of the outer horizon in $SL_2({\bf R})$ 
\footnote{If one takes the Kruskal coordinates, 
the horizon, not its analogue in $SL_2({\bf R})$, 
may be a two-dimensinal null surface.}. 
This can be read from the induced metric 
$ds^2_{(b_0,\tilde{b}_0)}$ of $Z_0$. 
The counterpart of the horizon is located at 
$\sigma=0$, that is, 
$\rho=\frac{1}{2}\ln \frac{b_0\tilde{b}_0}{(c/6)^2}$ 
and there the metric degenerates to  
\begin{eqnarray}
\left. ds^2_{(b_0,\tilde{b}_0)} 
\right|_{\rho=\frac{1}{2}\ln \frac{b_0\tilde{b}_0}{(c/6)^2}} 
=l^2 \left( \sqrt{\frac{b_0}{c/6}}dw-
\sqrt{\frac{\tilde{b}_0}{c/6}}d\tilde{w} \right)^2. 
\end{eqnarray}
As regards a generic $Z_{(b,\tilde{b})}$, 
the analogue of the horizon will be read from the singularity 
of the vector fields $\xi$ and $\tilde{\xi}$. 
Looking at (\ref{xi-barxi}), they diverge at 
$\rho=\frac{1}{2}\ln \frac{b\tilde{b}}{(c/6)^2}$. 
For the diffeomorphism to work, $\rho$ should  
satisfy $\rho > \frac{1}{2}\ln \frac{b_0\tilde{b}_0}{(c/6)^2}$. 
Only with this understanding 
$Z_{(b,\tilde{b})}$ becomes a deformation of $Z_0$. 
(The vector fields $\xi$ and $\tilde{\xi}$ can be expected to  
be integrated step by step from $Z_0$ to $Z_{(b,\tilde{b})}$ 
in $SL_2({\bf R})$.)
The completion of $Z_{(b,\tilde{b})}$ at 
$\rho=\frac{1}{2}\ln \frac{b_0\tilde{b}_0}{(c/6)^2}$ 
will be done by adding a two-dimensional space-like surface in 
$SL_2({\bf R})$. This can be also understood from 
the behavior of the induced metric  
\begin{eqnarray}
\left. 
ds^2_{(b,\tilde{b})} 
\right|_{\rho=\frac{1}{2}\ln \frac{b\tilde{b}}{(c/6)^2}} 
=l^2 \left( \sqrt{\frac{b}{c/6}}dw-
\sqrt{\frac{\tilde{b}}{c/6}}d\tilde{w} \right)^2 
+l^2 \left( 
\frac{b'}{4b}dw+\frac{\tilde{b}'}{4\tilde{b}}d\tilde{w}
\right)^2. 
\end{eqnarray} 
The difference under their completions in $SL_2({\bf R})$ 
clearly shows that 
$Z_{(b,\tilde{b})}$ is the different region from 
$Z_0$ as far as $b$ or $\tilde{b}$ is not constant.

        If one identifies $w$ and $\tilde{w}$ with 
the coordinates of circles, each deformation 
$\delta_fb$ and $\delta_{\tilde{f}}\tilde{b}$ 
given in (\ref{coadjoint action1}) can be regarded  
as the coadjoint action of the Virasoro algebra with the central charge $c$. 
Here $b$ and $\tilde{b}$ play the role 
of quadratic differentials on the circles. 
Let us consider the coadjoint orbits of $b_0$ and $\tilde{b}_0$ 
\footnote{Several terminologies used in this paragraph  
will be explained in the next section.
Here we hope to deliver our perspective in a simple manner.}.
We will call them respectively 
$W_{b_0}^{(+)}$ and $W_{\tilde{b}_0}^{(-)}$.
Their product $W_{b_0}^{(+)}\times W_{\tilde{b}_0}^{(-)}$ may be 
identified with the family of the regions $Z_{(b,\tilde{b})}$ 
of $SL_2({\bf R})$. 
Strictly speaking, 
we should consider not $Z_{(b,\tilde{b})}$ itself but 
its quotient obtained by the identification 
$(w,\tilde{w},\rho)\sim(w+2\pi,\tilde{w}-2\pi,\rho)$. 
This is because the information about $b_0$ and $\tilde{b}_0$, 
that is, 
about mass and angular momentum of the black hole,  
is encoded not in $Z_0$ itself but, 
through the parametrization (\ref{Z0 non-ex 2}), 
encoded in the way how to take the quotient.   
The actions of the Virasoro algebra are consistent with 
the equivalence relation used in the quotient. 
Therefore we can discuss about the Virasoro deformation 
of the quotient space.
Let $Y_{(b,\tilde{b})}$ be the quotient $Z_{(b,\tilde{b})}/\sim$. 
In particular 
$Y_{(b_0,\tilde{b}_0)}$  
is identified with $X_{(J,M)}^{+}$, the exterior of 
the outer horizon of the black hole $X_{(J,M)}$.   
The Virasoro groups 
$\widehat{{\it diff} S^1}_+ \times \widehat{{\it diff} S^1}_-$ 
deform the metric 
$ds^2_{X_{(J,M)}}=\pi_*ds^2_{(b_0,\tilde{b}_0)}$ of $X_{(J,M)}^+$ 
preserving its asymptotic form. 
The deformed metric $\pi_*ds^2_{(b,\tilde{b})}$ 
is nothing but 
the metric of $Y_{(b,\tilde{b})}$ which is 
related with $X_{(J,M)}^{+}$ by the coadjoint action. 
$SL_2({\bf R})$ itself plays the role of a classifying space 
in this argument. 
Namely the coadjoint action of 
$\widehat{{\it diff} S^1}_+ \times \widehat{{\it diff} S^1}_-$ 
deforms the region $Z_{0}$ without touching 
the metric of $SL_2({\bf R})$. 
The metric $ds^2_{(b,\tilde{b})}$ of each deformed region 
$Z_{(b,\tilde{b})}$ is the one induced from 
$ds^2_{SL_2({\bf R})}$ and determines  
the metric of $Y_{(b,\tilde{b})}$ by the projection $\pi$. 
It is the deformed metric of the black hole.

    It should be emphasized that the above Virasoro deformation 
can not transform the black hole $X^+_{(J,M)}$ to the black hole 
$X^+_{(J',M')}$ which mass $M'$ and angular momentum $J'$ are 
different from $M$ and $J$. This can be 
understood by the fact that any constant shifts of 
$b_0$ and $\tilde{b}_0$ are not allowed in (\ref{coadjoint action1}) 
since $f$ and $\tilde{f}$ are taken to be $2\pi$-periodic 
for the consistency with the equivalence relation used in the projection 
$\pi$. What the deformation generates are the oscillating modes of
$b(w)$ and $\tilde{b}(\tilde{w})$. The deformed metric 
$\pi_{*} ds^{2}_{(b,\tilde{b})}$ is not static since $w$ and
$\tilde{w}$ are originally the light cone coordinates (\ref{pi non-ex}).


\section{3-Geometries As The Virasoro Coadjoint Orbit}


\subsection{Coadjoint orbits of the Virasoro group}

       The Virasoro algebra is the Lie algebra of the Virasoro 
group $\widehat{{\it diff}S^1}$, which is the central extension of 
${\it diff}S^1$, the group of diffeomorphisms of a circle.  
It consists of 
vector fields $f(w)\frac{d}{dw}$ on the circle together with 
a central element $c$. 
A general element has the form $f(w)\frac{d}{dw}-iac$ 
with $a$ a real number. 
Let us write it by the pair $(f,a)$.  
The commutation relation is given by 
\begin{eqnarray}
\left[ (f,a_1),(g,a_2) \right]= 
\left(fg'-f'g,~
\frac{1}{48\pi}\! \int_0^{2\pi}\!dw(fg'''-f'''g)
\right).
\label{Virasoro 1}
\end{eqnarray} 
The coadjoint representation of the Virasoro algebra 
consists of quadratic differentials $b(w)(dw)^2$ together 
with the dual central element $c^*$. 
A general coadjoint vector 
(an element of the dual of the Virasoro algebra) 
has the form $b(w)(dw)^2+itc^*$ with $t$ a real number.  
Let us denote it by the pair $(b,t)$. 
The pairing between coadjoint  
and adjoint vectors are given by 
\begin{eqnarray}
\langle(b,t),(g,a)\rangle=
\frac{1}{2\pi}\int_0^{2\pi}dw b(w)g(w)~~+ta. 
\label{Virasoro dual pairing}
\end{eqnarray}
This is invariant under reparametrizations of the circle.
The coadjoint action of the Virasoro algebra is 
\begin{eqnarray}
\delta_f(b,t)
=(fb'+2f'b-\frac{t}{12}f''',0),
\label{coadjoint action 2}
\end{eqnarray}
where, the action of the central element being trivial, 
we write $\delta_{(f,a)}$ as $\delta_f$ for short.  
The dual pairing (\ref{Virasoro dual pairing}) 
is invariant under the action of the Virasoro algebra,
\begin{eqnarray} 
\delta_f\langle(b,t),(g,a)\rangle
&=&
\langle\delta_f(b,t),(g,a)\rangle+
\langle(b,t),\delta_f(g,a)\rangle 
\nonumber \\  
&=& 0, 
\end{eqnarray}
where $\delta_f(g,a)$ is given 
by (\ref{Virasoro 1}).

           As can be seen in (\ref{coadjoint action 2}) 
the dual central element $c^*$ is invariant under the 
coadjoint action. 
The value of $t$ determines the central charge. 
We shall set $t=\frac{3l}{2G}$. 
According to the definition (\ref{b0-tb0-c}) 
we will write this $t$ as $c$. 
And $W_{b_0}$, 
written as $W_{b_0, \tilde{b}_0}^{(\pm)}$ 
at the end of the previous section,  
is the coadjoint orbit of $(b_0,c)$, 
that is, the $\widehat{{\it diff} S^1}$-orbit of $(b_0,c)$.
The transformation (\ref{coadjoint action 2}) also shows 
that $W_{b_0}$ is a homogeneous space of ${\it diff}S^1$. 
Therefore we can parametrize the orbit by elements of 
${\it diff}S^1$ ( modulo the little group $H$ ) : 
For $s \in {\it diff}S^1$, letting $w_s=s(w)$, 
the integrated form of (\ref{coadjoint action1}) or 
(\ref{coadjoint action 2}) becomes 
\begin{eqnarray}
b^s(w_s)(dw_s)^2=
\left[ b_0+\frac{c}{12}\{w_s,w\} \right](dw)^2,  
\label{coadjoint action 3}
\end{eqnarray}
where $\{w_s,w \}$ is the Schwarzian derivative, 
$\{w_s,w\} \equiv$ 
$\frac{d^3 w_s/dw^3}{dw_s/dw}$
$-\frac{3}{2}\left( \frac{d^2w_s/dw^2}{dw_s/dw} \right)^2$.
Then $b^s$ provides the corresponding element of $W_{b_0}$.

          According to the Kirillov-Souriau-Kostant theory 
\cite{Woodhouse}, each coadjoint orbit admits to 
have a canonical symplectic structure defined by the dual pairing. 
If one regards tangent vectors at $b^s$ as vector fields on 
the circle,  
the symplectic form $\Omega$ can be described as follows : 
\begin{eqnarray}
\Omega|_{b^s}(u,v)=\langle (b^s,c),
\left[ (u,0),(v,0) \right] \rangle, 
\label{symplectic form}
\end{eqnarray} 
where $u=u(w_s),v=v(w_s)$ denote tangent vectors at $b^s$.

       Generators of the Virasoro algebra 
are given by Hamiltonian functions on $W_{b_0}$. 
If one takes $L_m=ie^{imw}\frac{d}{dw}$, 
the commutation relation (\ref{Virasoro 1}) becomes 
\begin{eqnarray}
\left[L_m,L_n \right] 
=(m-n)L_{m+n}+\frac{c}{12}m^3\delta_{m+n,0}.
\label{Virasoro cylinder}
\end{eqnarray}
The corresponding Hamiltonian functions are given by   
\begin{eqnarray}
l_m(s)=\langle (b^s,c),(e^{imw_s},0)\rangle, 
\label{lm(s)}
\end{eqnarray}   
where $s \in {\it diff}S^1$ is thought as the coordinate of 
$W_{b_0}$. 
Using the prescribed symplectic structure  \cite{Woodhouse} 
they turn out to satisfy the (classical) Virasoro algebra :  
\begin{eqnarray}
\{ l_m,l_n \}
=i(m-n)l_{m+n}+i \frac{c}{12}m^3\delta_{m+n,0}.
\label{classical Virasoro algebra}
\end{eqnarray}


\subsection{3-geometries as the Virasoro coadjoint orbit}

           We generalize our argument given 
in the previous section to the cases of 
the massive-extremal and massless black holes.   
For each of them there appears an appropriate region 
$Z_0$ of $SL_2({\bf R})$. The Virasoro groups 
$\widehat{{\it diff}S^1}_+\times \widehat{{\it diff}S^1}_-$ 
can deform each $Z_0$ in $SL_2({\bf R})$. 
This will be described by the coadjoint action of 
the Virasoro group. Quotient of the deformed 
region, obtained in a similar manner as the non-extremal case, 
provides a deformation of the exterior of the outer horizon. 
Each quotient has a metric induced from $SL_2({\bf R})$,  
which is nothing but the deformed metric of the black hole 
considered in \cite{Banados1}.    
The family of these quotients will be identified with the product of 
the coadjoint orbits 
$W_{b_0}^{(+)}\times W_{\tilde{b}_0}^{(-)}$. 
The values of $b_0$ and $\tilde{b}_{0}$ can be read from 
the values $(J,M)$ of the black hole by using (\ref{b0-tb0-c}). 
We also examine the case of $AdS_3$. In this case it becomes 
necessary to study $\widetilde{SL_2}({\bf R})$, the universal cover 
of $SL_2({\bf R})$. Taking an appropriate region of 
$\widetilde{SL_2}(\bf R)$, it becomes possible to repeat an argument 
similar to the black holes. In particular the family of the deformed 
regions will be identified with 
$W_{-c/24}^{(+)}\times W_{-c/24}^{(-)}$.

\subsubsection{Extremal black hole}

\underline{Massive case}  \\ 
Let $Z_0$ be a region of $SL_2({\bf R})$ 
consisting of the group elements
\footnote{
$J^{\pm}\equiv \pm J^0+J^2$.}
\begin{eqnarray}
e^{-\varphi J^2}
e^{\sigma J^1}
e^{\tilde{w} J^-}, 
\label{Z0 massive-ex}
\end{eqnarray}
where $\varphi,\tilde{w}$ and $\sigma$ are 
real parameters. There ranges are taken 
$-\infty < \varphi,\tilde{w},\sigma < +\infty$. 
It is a non-compact simply-connected region of 
$SL_2({\bf R})$. 
Regarding these parameters as coordinates of $Z_0$,   
the induced metric from $SL_2({\bf R})$ can be written 
in the form 
\begin{eqnarray}
ds^2_0=
\frac{l^2}{4}
\left \{(d\varphi)^2  
-2e^{\sigma}d\varphi d\tilde{w} 
+(d\sigma)^2 \right \}.
\label{ds0 massive-ex}
\end{eqnarray}
This non-compact simply-connected region of $SL_2({\bf R})$ 
is a covering space of $X_{(Ml,M)}^+$, the exterior of the 
outer horizon of the massive extremal black hole $X_{(Ml,M)}$. 
The projection $\pi$ can be described by  
\begin{eqnarray}
\frac{2r_+}{l}\left(\frac{t}{l}+\phi \right)
=\varphi,~~~
\frac{t}{l}-\phi =\tilde{w},~~~
\frac{r^2-r_+^2}{r_+l}=e^{\sigma},
\label{pi massive-ex}
\end{eqnarray}
with the identification $\phi \sim \phi+2\pi$. 
Since $\pi$ turns out to be the isometry,  
we can identify $X_{(Ml,M)}^+$ with the quotient 
of $Z_0$ :  
\begin{eqnarray}
X^+_{(Ml,M)}=Z_0/\sim.
\label{quotient massive-ex}
\end{eqnarray}
Here the equivalence relation is given by  
$(\varphi,\tilde{w},\sigma)\sim$  
$(\varphi+4\pi\frac{r_+}{l},\tilde{w}-2\pi,\sigma)$.

               Reparametrization of the group elements 
(\ref{Z0 massive-ex}) to the form  
\begin{eqnarray}
e^{-\sqrt{16GM}wJ^2}
e^{(\rho-\frac{1}{2}\ln4GM)J^1}
e^{\tilde{w}J^-} 
~~~~~~(\equiv h(w,\tilde{w},\rho)),
\label{h massive-ex 2}
\end{eqnarray}  
leads the flat $SL_2({\bf R})_+\times SL_2({\bf R})_-$ connection 
\begin{eqnarray}
A_0^{(+)}=
\left( \begin{array}{cc}
-\frac{1}{4}d\rho & -e^{\rho/2}dw \\
-4GMe^{-\rho/2}dw & \frac{1}{4}d\rho  
\end{array} \right),~~~
A_0^{(-)}=
\left( \begin{array}{cc}
\frac{1}{4}d\rho & 0 \\
-e^{\rho/2}d\tilde{w} & -\frac{1}{4}d\rho  
\end{array} \right).
\label{A0 massive-ex}
\end{eqnarray} 
In this derivation we first decompose 
the $SL_2({\bf R})$-valued function $h$ (\ref{h massive-ex 2}), 
according to the rule (\ref{def of hpm}),  
into the pair of $SL_2({\bf R})_{\pm}$-valued functions, 
$h^{(+)}=e^{-\frac{1}{2}(\rho-\ln 4GM)J^1}e^{\sqrt{16GM}wJ^2}$ 
and 
$h^{(-)}=e^{\frac{1}{2}\rho J^1}e^{\tilde{w}J^-}$, and then 
apply the prescription (\ref{def of A0}). 
The connection (\ref{A0 massive-ex}) coincides with  
(\ref{A(b,tb)}) at $(b,\tilde{b})=(Ml,0)$.

                      The equivalence relation 
used in the quotient (\ref{quotient massive-ex}) 
acquires the form (\ref{equiv non-ex}) in these new coordinates.  
So our argument completely reduces to  
that given in the previous section.    
In particular deformations of the massive-extremal 
black hole $X_{(Ml,M)}^+$ are realized by  
the quotients $Y_{(b,\tilde{b})}$ which are 
connected with it by the Virasoro group. 
The family of these $Y_{(b,\tilde{b})}$ can be identified 
with $W_{Ml}^{(+)} \times W_{0}^{(-)}$.

~

\noindent
\underline{Massless case}   \\ 
Let $Z_0$ be a region of $SL_2({\bf R})$ 
consisting of the group elements 
\begin{eqnarray}
e^{-w J^+}
e^{\rho J^1}
e^{\tilde{w} J^-}~~~~~
(\equiv h(w,\tilde{w},\rho)), 
\label{Z0 massless}
\end{eqnarray}
where $w,\tilde{w}$ and $\rho$ are real parameters. 
Their ranges are taken 
$-\infty < w,\tilde{w},\rho < +\infty$. 
It is also a non-compact simply-connected region 
of $SL_2({\bf R})$.  
Regarding these parameters as coordinates of $Z_0$, 
the induced metric from $SL_2({\bf R})$ can be written as  
\begin{eqnarray}
ds^2_0=
\frac{l^2}{4}
\left \{ 
-4e^{\rho}dw d\tilde{w} 
+(d\rho)^2 \right \}.
\label{ds0 massless}
\end{eqnarray}
This non-compact simply-connected region 
is a covering space of $X_{(0,0)}^+$, the exterior of 
the horizon of the massless black hole. 
The projection $\pi$ is prescribed by  
\begin{eqnarray}
\frac{t}{l}+\phi =w,~~~
\frac{t}{l}-\phi =\tilde{w},~~~
\frac{r}{l}=e^{\rho/2},
\label{pi massless}
\end{eqnarray}
with the identification $\phi \sim \phi+2\pi$.  
Hence the exterior of the horizon of 
the massless black hole can be identified with 
the quotient of $Z_0$ : 
\begin{eqnarray}
X^+_{(0,0)}=Z_0/\sim,
\label{quotient massless}
\end{eqnarray}
where the equivalence relation is 
$(w,\tilde{w},\rho)\sim$  
$(w+2\pi,\tilde{w}-2\pi,\rho)$.

               Decomposition of (\ref{Z0 massless}) 
by $h^{(+)}=e^{-\frac{1}{2}\rho J^1}
e^{wJ^+}$ and 
$h^{(-)}=e^{\frac{1}{2}\rho J^1}
e^{\tilde{w}J^-}$ 
leads the flat 
$SL_2({\bf R})_+\times SL_2({\bf R})_-$ connection 
\begin{eqnarray}
A_0^{(+)}=
\left( \begin{array}{cc}
-\frac{1}{4}d\rho & -e^{\rho/2}dw \\
0 & \frac{1}{4}d\rho  
\end{array} \right),~~~
A_0^{(-)}=
\left( \begin{array}{cc}
\frac{1}{4}d\rho & 0 \\
-e^{\rho/2}d\tilde{w} & -\frac{1}{4}d\rho  
\end{array} \right).
\label{A0 massless}
\end{eqnarray}
This is the connection (\ref{A(b,tb)}) 
at $(b,\tilde{b})=(0,0)$. So our argument also reduces to  
the previous one.    
Deformations of the massless black hole $X_{(0,0)}^+$ 
are realized by the quotients $Y_{(b,\tilde{b})}$ which are 
connected with it by the Virasoro group.   
The family of these $Y_{(b,\tilde{b})}$ can be identified 
with $W_{0}^{(+)} \times W_{0}^{(-)}$.


\subsubsection{Anti de-Sitter space}

       $AdS_3$ is the universal cover of $SL_2({\bf R})$. 
In terms of the Schwarzschild type coordinates 
$(t,\phi,r)$, where the ranges are 
$-\infty < t <+\infty$, $0 \leq \phi < 2\pi$ and 
$0 \leq r <+\infty$, the metric $ds_{AdS_3}^2$ acquires 
the form 
\begin{eqnarray}
ds_{AdS_3}^2=
-\left(1+\frac{r^2}{l^2}\right)^2(dt)^2+
\left(1+\frac{r^2}{l^2}\right)^{-2}(dr)^2+
r^2(d\phi)^2.
\label{ds AdS3}
\end{eqnarray}
Notice that $r=0$ is not a true singularity. 
It is merely a coordinate singularity.  
Nevertheless, to provide the Virasoro 
deformation of this space which is analogous 
to the black holes,  
it is required to consider the 
complement of $r=0$ rather than the whole. 
Actually this complement is an analogue of 
$X_{(J,M)}^+$ of the black hole. 
The complement of the line at $r=0$ 
will be called $AdS_3^+$. 

               Let $Z_0$ be a region of $SL_2({\bf R})$ 
consisting of the group elements 
\begin{eqnarray}
e^{-w J^0}
e^{\sigma J^1}
e^{-\tilde{w} J^0}, 
\label{Z0 ads}
\end{eqnarray}
where $w,\tilde{w}$ and $\sigma$ are real parameters. 
Notice that (\ref{Z0 ads}) have the form of the Cartan 
decomposition. In order to avoid a doubly 
parametrization, the ranges of $w$ and $\tilde{w}$ 
should be taken within $0\leq w \pm \tilde{w}<4\pi$. 
So we choose $0\leq w \pm \tilde{w}<4\pi$ for their ranges. 
As regards $\sigma$ it is set to $0<\sigma<+\infty$. 
If one admits $\sigma=0$ in the definition, 
$Z_0$ coincides with $SL_2({\bf R})$ itself.   
Due to its exclusion $Z_0$ becomes 
the complement of the circle at $\sigma=0$.  
If one regards the time of $AdS_3$ being $2\pi l$-periodic, 
$Z_0$ can be identified with $AdS_3^+$. 
The relation with the Schwarzschild type coordinates 
can be read as 
\begin{eqnarray}
w=\frac{t}{l}+\phi,~~~ 
\tilde{w}=\frac{t}{l}-\phi,~~~
e^{\sigma/2}=\sqrt{1+\frac{r^2}{l^2}}+\frac{r}{l}.
\end{eqnarray} 

    Introduce a slightly different parametrization of $Z_0$ 
with the form   
\begin{eqnarray}
e^{-w J^0}
e^{(\rho+2\ln2) J^1}
e^{-\tilde{w} J^0} 
~~~~~(\equiv h(w,\tilde{w},\rho)), 
\label{h ads 2}
\end{eqnarray}
where real parameter $\rho$ satisfies  
$-2\ln2 < \rho < +\infty$, due to the shift. 
Decomposition of the $SL_2({\bf R})$-valued function $h$ (\ref{h ads 2}) 
by the pair of $SL_2({\bf R})_{\pm}$-valued functions, 
$h^{(+)}=e^{-\frac{1}{2}(\rho+2\ln2)J^1}e^{wJ^0}$ 
and 
$h^{(-)}=e^{\frac{1}{2}(\rho+2\ln2)J^1}e^{-\tilde{w}J^0}$ 
leads the flat $SL_2({\bf R})_+\times SL_2({\bf R})_-$ connection 
\begin{eqnarray}
A_0^{(+)}=
\left( \begin{array}{cc}
-\frac{1}{4}d\rho & -e^{\rho/2}dw \\
\frac{1}{4}e^{-\rho/2}dw & \frac{1}{4}d\rho  
\end{array} \right),~~~
A_0^{(-)}=
\left( \begin{array}{cc}
\frac{1}{4}d\rho & \frac{1}{4}e^{-\rho/2}d\tilde{w} \\
-e^{\rho/2}d\tilde{w} & -\frac{1}{4}d\rho  
\end{array} \right).
\label{A0 ads}
\end{eqnarray}
This is the connection (\ref{A(b,tb)}) 
at $(b,\tilde{b})=(-c/24,-c/24)$.

$SL_2({\bf R})$ has a topologically nontrivial circle.
The universal cover $\widetilde{SL_2}({\bf R})$ 
can be obtained by making this $S^1$ to a line. 
This is equivalent to a simple change of 
the identification of $w$ and $\tilde{w}$ 
in the Cartan decomposition into an identification,   
$(w,\tilde{w})\sim(w+2\pi,\tilde{w}-2\pi)$. 
The universal covering of $Z_0$  
taken in $\widetilde{SL_2}({\bf R})$ is 
the complement of the line at $\sigma=0$ 
and can be identified with $AdS_3^+$.

For a given element $(b,\tilde{b})$ 
$\in W_{-c/24}^{(+)} \times W_{-c/24}^{(-)}$, 
the flat connection (\ref{A(b,tb)}) may be thought 
as a connection on a solid cylinder, 
where $\rho$ measures its radial 
direction and $(w,\tilde{w})$ are identified with 
the lightcone coordinates of the cylinder. 
Trivializations of $A^{(\pm)}_{b,\tilde{b}}$ 
and their local Lorentz invariant pairing 
define a map from the solid cylinder to $SL_2({\bf R})$. 
The solid cylinder wraps $SL_2({\bf R})$ infinitely 
many times. We may unfold this wrapping 
by considering $\widetilde{SL_2}({\bf R})$ in 
stead of $SL_2({\bf R})$. 
Therefore, letting $Z_0$ be the complement of the line in 
$\widetilde{SL_2}({\bf R})$ we can construct its deformation 
by the Virasoro group. 
The family of the deformations can be identified with 
$W_{-c/24}^{(+)} \times W_{-c/24}^{(-)}$.


\subsection{Quantization of the Virasoro coadjoint orbits}

For a given coadjoint orbit $W_{b_0}$,  
the Hamiltonian function $l_0(s)$ generates 
the $S^1$-action and can be considered as an energy 
function of the orbit. $b_0$ is a fixed point of this 
circle action and corresponds to the classical vacuum. 
We will discuss about the stability of this vacuum. 
Write $s \in \widehat{{\it diff}S^1}$ in the form 
$s(w)=w+\sum_ns_ne^{-inw}$, 
where $s_n$ are complex numbers satisfying $\bar{s}_n=s_{-n}$.   
These $s_n$ with $n \neq 0$ provide local coordinates 
in the neighborhood of $b_0$. 
$s_0$ acts trivially on $b_0$ (cf.(\ref{coadjoint action 2})). 
The behavior of the energy function in the neighborhood 
of $b_0$ can be seen by inserting the above $s$ into 
(\ref{lm(s)}) and expanding it with respect to $s_n$. 
The expansion turns out \cite{Witten2} to be the form 
\begin{eqnarray}
l_0(s)=
b_0+ 
\sum_{n}
n^2(b_0+\frac{c}{24}n^2)|s_n|^2+ 
O(s_n^3).
\end{eqnarray}
The stability of the classical vacuum will be assured by 
the condition, $l_0$ being bounded from below at $b_0$. 
It is achieved only when $b_0$ satisfies  
\begin{eqnarray}
b_0 \geq -\frac{c}{24}. 
\label{stability}
\end{eqnarray}
All the coadjoint orbits corresponding to 
the 3-geometries under consideration 
are satisfying this stability condition.

The orbit $W_{b_0}$ which satisfies the condition 
(\ref{stability}) can be quantized. 
This provides a unitary irreducible 
representation of the Virasoro algebra. 
Actually it is quantized 
\cite{Segal,B-R,Witten2} 
by the K\"ahler quantization 
or geometric quantization method. 
The coadjoint orbit is topologically the homogeneous 
space ${\it diff}S^1/H$. 
The little group $H$ is $S^1$ for the black holes, and 
$SL_2(\bf{R})$ for $AdS_3$ 
\footnote{Generators of $H$ can be seen easily by solving 
$\delta_f(b_0,c)=0$ (\ref{coadjoint action 2}) for each cases.}.
For these $H$, the homogeneous space ${\it diff}S^1/H$ becomes 
a complex manifold. 
The complex structure turns out \cite{Witten2} to be compatible 
with the symplectic structure (\ref{symplectic form}). 
Thereby the orbit $W_{b_0}$ which we need, becomes K\"ahler. 
The complex line bundle on the orbit with its $1$st Chern 
class $\Omega$ provides the unitary irreducible representation 
on the space of the holomorphic sections.

      To describe the representations it is convenient to shift the 
Virasoro generator $L_0$ to $L_0+\frac{c}{24}$. 
With this shift the algebra (\ref{Virasoro cylinder}) becomes 
the standard form 
\begin{eqnarray}
\left[L_m,L_n \right] 
=(m-n)L_{m+n}+\frac{c}{12}(m^3-m)\delta_{m+n,0}.
\label{Virasoro plane}
\end{eqnarray}
Any representation of the Virasoro algebra with the central charge 
$c$ can be specified by its highest weight state $|h \rangle$.  
The state $|h\rangle$, which is called primary,   
satisfies $L_0|h \rangle= h| h \rangle$ and   
$L_n |h \rangle=0$ for $\forall n \geq 1$.  
In addition to these conditions 
the primary state $|0\rangle$ satisfies 
$L_{-1}|0 \rangle =0$. 
So it is a $SL_2({\bf R})$ invariant state.

             Unitary representations are the representations 
in which $L_n$ satisfy the condition 
$L_n^{\dagger}=L_{-n}$. The unitary 
irreducible representations which are obtained by 
the quantization of the orbits 
can be summarized as follows \cite{Witten2}
\footnote{Here the condition $c \geq 1$ is required.}: 
For the orbit of $W_{b_0}$ with $b_0 > -c/24$, 
the corresponding unitary irreducible 
representation is given by the Verma 
module ${\cal V}_{h=b_0+\frac{c}{24}}$. 
It is a module obtained by
successive actions of $L_{-n}$ ($n\geq 1$) 
on $|b_0+\frac{c}{24}\rangle$. 
It has the form 
\begin{eqnarray}
{\cal V}_{b_0+\frac{c}{24}}
= \bigoplus_{n_1 \geq n_2 \geq \cdots \geq n_k \geq 1}
{\bf C}L_{-n_1}L_{-n_2}\cdots L_{-n_k}
|b_0+\frac{c}{24} \rangle. 
\label{b0>-c/24}
\end{eqnarray} 
For the orbit $W_{-c/24}$, 
the corresponding unitary irreducible 
representation is given by an analogue of the Verma module 
${\cal V}_0$.   
It is obtained by successive actions of 
$L_{-n}$ ($n \geq 2$) on $|0 \rangle$. 
We call this module ${\cal V}'_0$. It has the form  
\begin{eqnarray}
{\cal V}'_0= \bigoplus_{n_1 \geq n_2 \geq \cdots \geq n_k \geq 2}
{\bf C}L_{-n_1}L_{-n_2}\cdots L_{-n_k}
|0 \rangle. 
\label{b0=-c/24}
\end{eqnarray}

~

                Gathering these results about 
quantization of the orbits 
we can prescribe quantization of 
the asymptotic Virasoro symmetry of the 3-geometries 
or their Virasoro deformations in the following manner : 
For the BTZ black hole $X_{(J,M)}$, the deformations 
$Y_{(b,\tilde{b})}$ of the exterior of the outer horizon  
can be identified with the product of the coadjoint orbits 
$W_{b_0}^{(+)}\times W_{\tilde{b}_0}^{(-)}$.  
The quantizations of the orbits 
$W_{b_0}^{(+)}$ and $W_{\tilde{b}_0}^{(-)}$ 
provide the unitary irreducible representations 
${\cal V}^{(+)}_{b_0+\frac{c}{24}}$ and 
${\cal V}^{(-)}_{b_0+\frac{c}{24}}$. 
Therefore the quantization of the deformations  
leads the representation  
${\cal V}^{(+)}_{b_0+\frac{c}{24}}
\times {\cal V}^{(-)}_{\tilde{b}_0+\frac{c}{24}}$. 
In particular state of the black hole can be 
identified with the primary state  
$|b_0+\frac{c}{24} \rangle \otimes$ 
$|\tilde{b}_0+\frac{c}{24} \rangle $ 
of the representation . 
For $AdS_3$, 
the deformations of $AdS_3^+$ can be identified with 
the product of the coadjoint orbits  
$W_{-\frac{c}{24}}^{(+)}\times 
W_{-\frac{c}{24}}^{(-)}$.  
These orbits provide the unitary 
irreducible representations  
${{\cal V}'_{0}}^{(+)}$ and ${{\cal V}'_{0}}^{(-)}$.  
Therefore the quantization of the deformation 
leads the representation 
${{\cal V}'_{0}}^{(+)}
\times {{\cal V}'_{0}}^{(-)}$. 
The state of $AdS_3$ can be 
identified with the primary state  
$|0 \rangle \otimes$ 
$|0 \rangle $.  
It is the state invariant 
under $SL_2({\bf R})_+ \times SL_2({\bf R})_-$,  
which is the isometry of $AdS_3$.

    Excitations by $L_{-n}$ correspond to the Virasoro deformation 
of $Y_{(b_0,\tilde{b}_0)}=Z_0/\sim$. 
Originally it is the deformation of $Z_0$ in $SL_2({\bf R})$ 
which by no means provides any transformation of the black hole 
to another one having different mass and angular momentum 
\footnote{It is not a reparametrization of $Z_0$.}. 
Nevertheless, these degrees of freedom provide 
the unitary representations of the Virasoro algebra. 
This shows that these degrees of freedom can be included 
in the physical spectrum as (massive) gravitons 
\cite{F-K-M}.


\section{Quantization Of Three-Dimensional Gravity}

As we have seen, 
quantizations of the Virasoro deformations of 
the BTZ black holes and $AdS_3$ lead the unitary irreducible 
representations of the Virasoro algebra at least when 
$\frac{3l}{2G}\geq 1$.  One may wonder quantization of  
three-dimensional gravity with negative cosmological 
constant becomes complete by these quantizations. 
But it does not so. 
If one takes the Chern-Simons gravity viewpoint, 
these deformations correspond to the local degrees 
of freedom of the theory (gravitons), 
that is, oscillating modes of $b$ and $\tilde{b}$ 
in the flat connection (\ref{A(b,tb)}).  
To make the quantization  
\footnote{It would give a quantum gravity 
in the non-topological phase \cite{Witten1}.}
complete, 
one must take into account of the global degrees 
of freedom, {\it i.e.}, holonomies. 
The holonomy variables should be dynamical in the 
Chern-Simons gauge theory. In our context 
this requires an introduction of their conjugates.  
Given a suitable Poisson structure on the holonomy 
variables and their quantizations, 
it is very reasonable to expect from the perspective 
of the $AdS_3/CFT_2$ correspondence that, 
with an identification of these quantum operators 
with the zero modes of an appropriate 
two-dimensional quantum field, 
the unitary representations obtained in the previous 
section could be reproduced as the Hilbert space of 
the two-dimensional conformal field theory.  
This expectation turns out to be true at least when 
$\frac{3l}{2G}\gg 1$. The holonomy variables can be 
identified with the zero modes of a real scalar field $X$.
This scalar field should be interpreted as the Liouville 
field  in the ultimate.

            We start with the description of 
the Poisson structure of the holonomies. 
Then we turn to construction of 
the unitary irreducible representations 
in the framework of the Liouville field theory 
or the Coulomb gas formalism.  
Finally we make identifications 
between the states of the three-dimensional 
gravity and those of the Liouville field theory.


\subsection{Holonomy variable}

                For a given closed path $\gamma$, 
holonomy of a connection $A=(A^{(+)},A^{(-)})$ along the path 
is given by   
\begin{equation}
H^{(\pm)}[\gamma] \equiv Pe^{\int_\gamma A^{(\pm)}},
\end{equation}
where $P$ denotes the path ordering. 
Since we can assume $A$ is flat,  
$H[\gamma]=(H^{(+)}[\gamma], H^{(-)}[\gamma])$ is 
invariant up to conjugation under smooth deformations of $\gamma$.
In particular ${\rm Tr}H^{(\pm)}$ are homotopy invariant.
The BTZ black holes originally admit to have a nontrivial holonomy 
around a closed path connecting $(w, \tilde{w}, \rho)$ and 
$(w+2\pi, \tilde{w}-2\pi, \rho)$ 
in the constant $\rho$ surface.  
Let us call this path $\gamma_1$. 
If there is only one nontrivial holonomy 
there appears no symplectic structure. 
In order to make it a dynamical variable 
we need to introduce another ``closed" path. 
Let $\gamma_2$ be a path connecting 
$(w, \tilde{w}, \rho)$ and 
$(w+\sqrt{\frac{c/6}{b_0}}\Sigma, \tilde{w}
+\sqrt{\frac{c/6}{\tilde{b}_0}}\tilde{\Sigma}, \rho)$ 
in the constant $\rho$ surface. 
If one regards it as a closed path 
and then considers about a holonomy around this path, 
the quantities $\Sigma$ and $\tilde{\Sigma}$ 
together with $r_{\pm}$ may become dynamical. 
In this extended ``phase" space, the BTZ black holes 
themselves will become a Lagrangian submanifold 
$\Sigma=\tilde{\Sigma}=0$. 
This extension is a generalization of the off-shell 
extension of the Euclidean black holes examined by 
Carlip and Teitelboim \cite{Carlip-Teitelboim}.

The flat connection  
which describes $X^+_{(J,M)}$, 
the exterior of the outer horizon 
of the non-extremal BTZ black hole,  
is given by (\ref{A0 non-ex}). The holonomy of this connection 
around the path $\gamma_1$ can be evaluated as   
\begin{eqnarray}
H^{(+)}[\gamma_1] 
&=& h^{(+)}(w+2\pi, \tilde{w}-2\pi, \rho)
   {h^{(+)}}^{-1}(w, \tilde{w}, \rho) \nonumber \\
&=& e^{-\frac12(\rho-\ln\frac{b_0}{c/6})J^1} 
      e^{2\pi \sqrt{\frac{b_0}{c/24}} J^2}
       e^{\frac12(\rho-\ln\frac{b_0}{c/6})J^1}, \nonumber \\
H^{(-)}[\gamma_1] 
&=& h^{(-)}(w+2\pi, \tilde{w}-2\pi, \rho)
   {h^{(-)}}^{-1}(w, \tilde{w}, \rho) \nonumber \\
&=& e^{\frac12(\rho-\ln\frac{\tilde{b}_0}{c/6})J^1}
      e^{-2\pi \sqrt{\frac{\tilde{b}_0}{c/24}}J^2}
       e^{-\frac12(\rho-\ln\frac{\tilde{b}_0}{c/6})J^1}, 
\label{holonomy 1}
\end{eqnarray}
where $h^{(\pm)}$ are those used in its trivialization 
and their explicit forms are given in (\ref{hpm non-ex}). 
The trace becomes 
\begin{eqnarray}
{\rm Tr}H^{(+)}[\gamma_1] 
= 2\cosh\pi\sqrt{\frac{b_0}{c/24}},
~~~
{\rm Tr}H^{(-)}[\gamma_1] 
= 2\cosh\pi\sqrt{\frac{\tilde{b}_0}{c/24}}.
\label{trace 1}
\end{eqnarray}
Similarly its holonomy around the path $\gamma_2$ has the form  
\begin{eqnarray}
H^{(+)}[\gamma_2] &=& 
     e^{-\frac12(\rho-\ln\frac{b_0}{c/6})J^1} e^{2\Sigma J^2}
      e^{\frac12(\rho-\ln\frac{b_0}{c/6})J^1}, \nonumber \\
H^{(-)}[\gamma_2] &=& 
     e^{\frac12(\rho-\ln\frac{\tilde{b}_0}{c/6})J^1}
      e^{2\tilde{\Sigma}J^2}
       e^{-\frac12(\rho-\ln\frac{\tilde{b}_0}{c/6})J^1}.
\label{holonomy 2}
\end{eqnarray}
Therefore we obtain 
\begin{eqnarray}
{\rm Tr}H^{(+)}[\gamma_2] 
= 2\cosh\Sigma,~~~
{\rm Tr}H^{(-)}[\gamma_2] 
= 2\cosh\tilde{\Sigma}.
\label{trace 2}
\end{eqnarray}

      To describe the Poisson structure of 
these holonomies in the Chern-Simons gravity,  
we follow the recipe developed by Nelson, Regge 
and  Zertuche  \cite{N-R-Z}.  
It states that holonomies in a constant 
time surface have non-vanishing Poisson brackets 
when the underlying paths intersect with each other. 
If one regards the radial coordinate $\rho$ as a time,  
the following Poisson algebra 
of $\mbox{Tr}H^{(\pm)}[\gamma_i] (i=1, 2)$ 
can be obtained :
\begin{eqnarray}
\left\{
{\rm Tr}H^{(\pm)}[\gamma_1], 
{\rm Tr}H^{(\pm)}[\gamma_2] 
\right\}
=\mp\frac{6\pi}{c} 
\left[
  -\left({\rm Tr}H^{(\pm)}[\gamma_1] \right)
     \left({\rm Tr}H^{(\pm)}[\gamma_2] \right)
    +2{\rm Tr}H^{(\pm)}[\gamma_2\gamma_1] \right].
\label{Poisson 1}
\end{eqnarray}
Substitution of the explicit forms 
(\ref{trace 1}) and (\ref{trace 2}) simplifies the expression 
(\ref{Poisson 1}) to  
\begin{eqnarray}
\left \{ 
\cosh\pi \frac{r_+ +r_-}{l}, \cosh\Sigma 
\right \} 
&=& 
-\frac{4\pi G}{l} 
\sinh\pi \frac{r_+ +r_-}{l} \sinh\Sigma, 
\nonumber \\
\left\{ 
\cosh\pi \frac{r_+ -r_-}{l}, 
\cosh \tilde{\Sigma} \right\} 
&=& 
 -\frac{4\pi G}{l} 
\sinh\pi \frac{r_+ -r_-}{l} 
\sinh \tilde{\Sigma}.
\label{Poisson 2}
\end{eqnarray}
These indicate that 
$r_+, r_-, \Sigma$ and $\tilde{\Sigma}$ 
become canonical variables with the Poisson algebra,
\begin{eqnarray}
\left\{ r_+ +r_-, \Sigma \right\} = -4G,~~~
\left\{ r_+ -r_-, \tilde{\Sigma} \right\} = -4G.
\label{Poisson 3}
\end{eqnarray}
Having obtained the symplectic structure,  
one can quantize these variables. 
Let $\hat{r}_+, \hat{r}_-, \hat{\Sigma}$ 
and $\hat{\tilde{\Sigma}}$ 
be the corresponding operators. 
Their nontrivial commutation relations are  
\begin{eqnarray}
\Big[\hat{r}_+ +\hat{r}_-, \hat{\Sigma}\Big] 
= -4iG,~~~~ 
\Big[\hat{r}_+ -\hat{r}_-, \hat{\tilde{\Sigma}}\Big] 
= -4iG. 
\label{CCR holonomy}
\end{eqnarray}


\subsection{Realization by the Liouville field theory}

     Let $X$ be a real scalar field on ${\bf P}^1$. 
\footnote{To use the conventional technique of $2d$ 
CFT we consider the Euclidean version. } 
The action is given by  
\begin{eqnarray}
S[X]=\frac1{4\pi i}\int_{{\bf P}^1} 
\bar{\partial} X \wedge \partial X
     + \frac{\alpha_0}{2\pi}\int_{{\bf P}^1} RX,
\label{Liouville action}
\end{eqnarray}
where $R$ is the Riemann tensor of 
a fixed K\"ahler metric on ${\bf P}^1$. 
\footnote{ 
$\frac{1}{2\pi} \int_{{\bf P}^1} R
=\chi({\bf P}^1)=2$.}
$\alpha_0$ is a real number and its value 
will be specified later.

           If one takes the following expansion 
of $X$ at $z=0$ 
\footnote{For simplicity we will only describe 
the holomorphic part of the theory. 
The operator product expansion of $X(z)$ 
(holomorphic part of $X(z,\bar{z})$) becomes 
$ X(z)X(w)\sim-\ln(z-w) $.}  
\begin{equation}
X(z)=x-ip\ln z+i\sum_{n\neq 0}\frac{a_n}n z^{-n},
\end{equation}
the nontrivial commutation relations 
among the mode operators become  
\begin{eqnarray}
\left[x, p \right] = i,~~~  
\left[a_m, a_n \right] = m \delta_{m+n,0},  
\end{eqnarray}
where $a_0 \equiv p$. 
Using these operators we will define 
the in-Fock vacuum $|k \rangle_{in}$ ($k \in {\bf C}$) 
in a following manner : 
Let $|0\rangle_{in}$ be the state defined by the conditions
\begin{eqnarray}
a_n|0 \rangle_{in} 
= 0 \hspace{20mm} 
{\rm for}~~ \forall n \geq 0.
\label{in 0-vacua}
\end{eqnarray}
Then $|k \rangle_{in}$ is introduced as the state 
obtained from $|0 \rangle_{in}$ by the relation 
\begin{eqnarray}
|k \rangle_{in} 
= e^{ikx}|0 \rangle. 
\label{in k-vacua}
\end{eqnarray}  
It satisfies 
$p|k \rangle_{in}=k|k \rangle_{in}$. 
The in-Fock space ${\cal F}_k^{in}$ is the Fock space 
built on $|k \rangle_{in}$.  
It has the form,  
$\displaystyle{
{\cal F}_k^{in}=    
\bigoplus_{n_1 \geq n_2 \geq \cdots \geq n_p \geq 1}
{\bf C}
a_{-n_1}a_{-n_2}\cdots a_{-n_p}|k \rangle_{in}}$. 
This space is located at $z=0$.

               The stress tensor $T$ has the form
\begin{eqnarray}
T(z)=-\frac{1}{2} 
\partial X \bar{\partial}X(z)+\alpha_0\partial^2 X(z). 
\label{stress tensor}
\end{eqnarray}
Expansion of the stress tensor to the form 
$T(z)=\sum_{n}L_nz^{-n-2}$ provides the generators 
of the Virasoro algebra (\ref{Virasoro plane}) 
with the central charge $1+12\alpha_0^2$. 
In terms of the oscillator modes their expressions become     
\begin{eqnarray}
L_0 &=& \sum_{k\geq1}a_{-k}a_k+\frac{1}{2}
p^2+i\alpha_0 p, 
\nonumber \\ 
L_n &=& \frac{1}{2} 
\sum_{k}a_{-k}a_{k+n}+i(n+1)\alpha_0 a_n 
~~~~~ {\rm for}  ~\forall  n \neq 0. 
\label{Ln by an}
\end{eqnarray}
The actions of $L_n$ on the in-Fock vacua satisfy the 
following properties :  
On the vacuum $|0\rangle_{in}$, 
using the expressions (\ref{Ln by an}) one can read 
\begin{eqnarray}
L_n |0 \rangle_{in}=0 \hspace{20mm} 
{\rm for}~\forall n \geq -1, 
\label{Ln on in-vacuum 0}
\end{eqnarray} 
while on the vacuum $|k \rangle_{in}$ 
with $k \neq 0$, 
one obtains  
\begin{eqnarray} 
L_0|k\rangle_{in} 
&=& 
\left\{\frac{(k+i\alpha_0)^2}2 
          +\frac{\alpha_0^2}2 \right\}|k\rangle_{in},
\nonumber \\ 
L_n |k \rangle_{in}
&=&
0 \hspace{20mm} 
\mbox{for}~\forall n \geq 1. 
\label{Ln on in-vacuum k}
\end{eqnarray}
These properties 
besides a comparison between the commutation relations 
$\left[L_0, a_{-n} \right]=na_{-n}$ and 
$\left[L_0, L_{-n} \right]=nL_{-n}$, 
show that the in-Fock space 
${\cal F}^{in}_k$ 
is equivalent to the Verma module 
${\cal V}_{h=\frac{1}{2}(k+i\alpha_0)^2
+\frac{1}{2}\alpha_0^2}$ 
with $c=1+12\alpha_0^2$,  
or ${\cal V}'_0$ if $k=0$.

            Hermitian conjugation of the mode 
operators becomes 
\begin{eqnarray}
a_n^{\dagger} &=& a_{-n}  \hspace{20mm}
{\rm for}~\forall  n \neq 0,  \nonumber \\
p^{\dagger} &=& p+2i\alpha_0, \nonumber \\
x^{\dagger} &=& x.  
\label{conjugate of modes}
\end{eqnarray}
Notice that $p$ can not be 
hermitian  due to the existence of 
the background charge $\alpha_0$.
Since the action (\ref{Liouville action}) 
is real, it is expected that the realization  
(\ref{Ln by an}) of the Virasoro algebra  
becomes unitary. In fact, 
under this conjugation the realization becomes unitary, 
that is, $L_n$ satisfy $L_n^{\dagger}=L_{-n}$.  
The unitarity imposes a constraint on the allowed value  
$k$ of the in-Fock vacuum. 
The condition $L_0^{\dagger}=L_0$ 
together with the requirement 
of its eigenvalue being non-negative, 
restricts $|k\rangle_{in}$ in the form 
(Fig.4)
\begin{equation}
k=\beta-i\alpha_0  \hspace{20mm} \beta\in {\bf R}_{\geq0},  
\label{k for black holes}
\end{equation}
or
\begin{equation}
k=i(\gamma-1)\alpha_0  \hspace{20mm} \gamma\in (0, 1]. 
\label{k for ads}
\end{equation}

\begin{figure}[t]
\epsfysize=5cm
\makebox[17cm][c]{\epsfbox{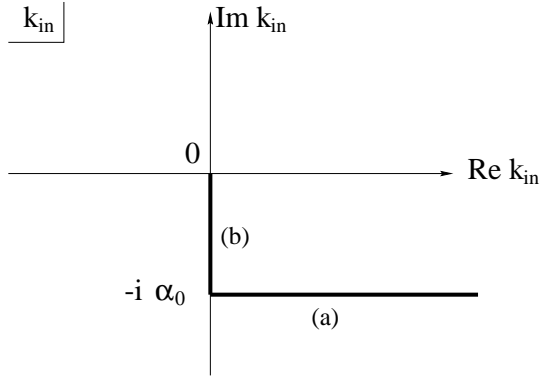}}
\caption{{\small The in-momentum allowed by the unitarity constraint.}}
\end{figure}
              
            Now we introduce the out-Fock space 
${\cal F}_{k'}^{out}$.    
Let $_{out}\langle 0|$ be the state which satisfies 
the conditions : 
$_{out}\langle 0|a_n^{\dagger} = 0 
~({\rm for}~ \forall  n \geq 0)$.
The out-Fock vacuum $_{out}\langle k'|$ can be introduced by 
$_{out}\langle k'| = _{out}\langle 0|e^{-ik'x}$. 
It satisfies $_{out}\langle k'|p= 
{}_{out}\langle k'|(k'-2i\alpha_0)$.
The out-Fock space ${\cal F}_{k'}^{out}$ is the Fock space 
built on $_{out}\langle k'|$. 
Using (\ref{conjugate of modes}) it has the form, 
$\displaystyle{{\cal F}_{k'}^{out}=    
\bigoplus_{n_1 \geq n_2 \geq \cdots \geq n_p \geq 1}{\bf C}
{}_{out} \langle k'|a_{n_p}\cdots a_{n_2}a_{n_1}}$. 
This space is located at $z=\infty$.

      The actions of $L_n$  
on the out-Fock vacua satisfy the following properties : 
On the vacuum $_{out}\langle 0|$, using the expressions 
(\ref{Ln by an}) one can obtain  
\begin{eqnarray}
{}_{out}\langle 0| L_{-n} =0 \hspace{20mm} 
{\rm for}~\forall n \geq -1, 
\label{Ln on out-vacuum 0}
\end{eqnarray} 
while on the vacuum $_{out}\langle k'|$ with $k' \neq 0$,  
\begin{eqnarray} 
{}_{out}\langle k'|L_0 &=&  
{}_{out}\langle k'|
  \left\{\frac{(k'-i\alpha_0)^2}2 
          +\frac{\alpha_0^2}2 \right\},
\nonumber \\ 
{}_{out}\langle k'| L_{-n} &=&0 \hspace{20mm} 
{\rm for}~\forall n \geq 1. 
\label{Ln on out-vacuum k'}
\end{eqnarray}
These properties besides the action of $p$ on the out-Fock vacuum 
show that the out-Fock space ${\cal F}^{out}_{k'=k+2i\alpha_0}$ 
can be identified with ${}^{*}{\cal F}^{in}_{k}$, the dual of 
${\cal F}^{in}_k$
\footnote{
We take the convention so that  
the state ${}^{*}_{in}\langle k|$, 
which is dual to $|k \rangle_{in}$, satisfies 
${}^{*}_{in}\langle k|p={}^{*}_{in}\langle k|k$. }. 
With this identification the pairing between 
${\cal F}^{out}_{k+2i\alpha_0}$ and ${\cal F}^{in}_k$ becomes 
non-degenerate. In particular, if one takes the allowed value 
(\ref{k for black holes}) or (\ref{k for ads}) of $k$,  
the representation becomes unitary. Choosing the value of 
$\alpha_0$ appropriately,  
it becomes the unitary irreducible representation 
of the Virasoro algebra 
obtained by the quantization of the coadjoint orbit.


\subsection{Quantization of $3d$ gravity 
based on the Liouville field theory}

                       Given the commutation relations 
(\ref{CCR holonomy}) of the quantum holonomy operators  
$\hat{r}_+, \hat{r}_-, \hat{\Sigma}$ 
and $\hat{\tilde{\Sigma}}$, 
one is tempted to identify these topological operators 
with the zero modes of the Liouville field $X(z,\bar{z})$. 
Let $x~(\tilde{x})$ and  $p~(\tilde{p})$ 
be the zero modes of the holomorphic part $X(z)$ 
(the anti-holomorphic part $X(\bar{z})$) of the Liouville 
field. The identification is precisely given by   
\begin{eqnarray}
x=\sqrt{\frac l{2G}}\hat{\Sigma},&&  
\tilde{x}=\sqrt{\frac l{2G}}\hat{\tilde{\Sigma}}, 
\nonumber \\
p=\frac{\hat{r}_+ +\hat{r}_-}{\sqrt{8Gl}}, && 
\tilde{p}=\frac{\hat{r}_+ -\hat{r}_-}{\sqrt{8Gl}},  
\label{identify zero modes}
\end{eqnarray}
and becomes consistent with 
the commutation relations $[x, p]=i$ 
and $[\tilde{x}, \tilde{p}]=i$.

\subsubsection{Identification of states}

           At the end of the previous section,  
the state of the BTZ black hole $X_{(J,M)}^+$ 
has been identified with the primary state 
$|b_0+\frac{c}{24} \rangle \otimes$ 
$|\tilde{b}_0+\frac{c}{24} \rangle $.   
The values of weights can be expressed in 
terms of the geometrical data,    
\begin{eqnarray}
b_0+\frac{c}{24}
&=& 
\frac{1}{16Gl}(r_++r_-)^2+\frac{l}{16G}, 
\nonumber \\ 
\tilde{b}_0+\frac{c}{24}
&=&
\frac{1}{16Gl}(r_+-r_-)^2+\frac{l}{16G}. 
\label{wt of X(J,M)}
\end{eqnarray}
A comparison of these weights with that 
of the in-Fock vacuum   
given in (\ref{Ln on in-vacuum k}) 
makes us adjust the background charge 
of the Liouville field to 
\begin{eqnarray}
\alpha_0=\sqrt{\frac{l}{8G}},    
\label{value of background charge}
\end{eqnarray}
and, accepting this background charge, 
it also leads an identification 
of the black hole state 
with the following in-Fock vacuum :  
\begin{equation}
X^+_{(J, M)} \Longleftrightarrow |J, M\rangle_{in} \equiv 
  |k_{(J, M)}\rangle_{in} \otimes |\tilde{k}_{(J, M)}\rangle_{in}, 
\label{black hole state in 2d}
\end{equation}
where $k_{(J,M)}$ and $\tilde{k}_{(J,M)}$ are given by 
\begin{eqnarray}
k_{(J, M)} &\equiv& 
-i\sqrt{\frac l{8G}}+\frac{r_+ +r_-}{\sqrt{8Gl}},
\nonumber \\ 
\tilde{k}_{(J, M)} &\equiv& 
 -i\sqrt{\frac l{8G}}+\frac{r_+ -r_-}{\sqrt{8Gl}}.   
\label{k(J,M)}
\end{eqnarray}    
These are the values allowed 
by the unitarity condition (\ref{k for black holes}). 
The region $(a)$ in Fig.4 corresponds to the black holes. 
All the excitations in the Fock space 
${\cal F}_{k_{(J,M)}}^{in} \times 
\overline{{\cal F}}^{in}_{\tilde{k}_{(J,M)}}$ can be 
identified with the excitations owing to the Virasoro 
deformation of $X_{(J,M)}^+$. These correspond to gravitons.

           To work the above machinery completely,  
the central charge (\ref{b0-tb0-c})
of the Virasoro deformation of the 3-geometries 
should be matched with 
that of the two-dimensional conformal field theory. 
This consistency requires the condition,  
$c \gg 1$ or equivalently 
$\sqrt{\frac{l}{8G}}\gg 1$. 
Under this condition 
the central charge of the Liouville field theory can 
be regarded as 
$1+12\alpha_0^2=1+\frac{3l}{2G}\sim \frac{3l}{2G}$. 
Therefore these two coincide in this limit. 
The deviation at finite $c$ is of order $1/c$.

               The state of $AdS_3^+$ has been 
identified with the primary state 
$|0\rangle \otimes |0\rangle$. 
It is invariant under 
$SL_2({\bf R})_+ \times SL_2({\bf R})_-$. 
Thereby we can identify this state with 
the $sl_2({\bf C})$-invariant vacuum of the Liouville 
field theory :  
\begin{equation}
AdS_3 \Longleftrightarrow |vac\rangle_{in}
\equiv |0\rangle_{in}\otimes|0\rangle_{in}.
\label{ads state in 2d}
\end{equation}

              The $sl_2({\bf C})$-invariant vacuum 
is the Fock vacuum with $\gamma=1$ 
($\tilde{\gamma}=1$) in (\ref{k for ads}) 
(the origin in Fig.4). 
From the viewpoint of the unitary representation theory 
other values of $\gamma$ and $\tilde{\gamma}$  
are also allowed as far as they satisfy 
$0 < \gamma, \tilde{\gamma} \leq 1$. Notice that 
$(\gamma,\tilde{\gamma})=(0,0)$, that is, 
$(\beta,\tilde{\beta})=(0,0)$ in (\ref{k for black holes}) 
corresponds to the state of the massless black hole 
$X_{(0,0)}^+$.   
Consider a state 
$|k_{\gamma}\rangle_{in} \otimes 
|\tilde{k}_{\tilde{\gamma}}\rangle_{in}$, 
where $k_{\gamma}$ and $\tilde{k}_{\tilde{\gamma}}$ 
have the form given in (\ref{k for ads}). 
The weights of this state are respectively  
$-\frac{\gamma^2 l}{16G}+\frac{l}{16G}$ and 
$-\frac{\tilde{\gamma}^2 l}{16G}+\frac{l}{16G}$. 
Using (\ref{wt of X(J,M)}) 
one can read mass and angular momentum 
as $M=-\frac{\gamma^2+\tilde{\gamma}^2}{16G}$ 
and $J=-\frac{(\gamma^2-\tilde{\gamma}^2)l}{16G}$. 
Therefore the corresponding 3-geometry 
is conic \cite{D-J}. 
This means that the state 
$|k_{\gamma}\rangle_{in} \otimes 
|\tilde{k}_{\tilde{\gamma}}\rangle_{in}$ 
is that of a conical singularity. 
The region $(b)$ in Fig.4 corresponds to 
the conical singularities.     
Rather surprisingly, its Virasoro deformation 
gives rise to the unitary representation. 
This might imply that conical singularities 
with the allowed mass and momentum get mild 
quantum mechanically. Our understanding of singularities 
in classical relativity may be required 
to change in quantum theory.

     The values of $b_0$ and $\tilde{b}_0$ which 
correspond to $k_{\gamma}$ and $\tilde{k}_{\tilde{\gamma}}$ 
are $-\frac{\gamma^2l}{16G}$ and 
$-\frac{\tilde{\gamma}^2l}{16G}$. 
The three-dimensional metric becomes 
\begin{eqnarray}
&&ds^2_{(b_0,\tilde{b}_0)=(-\frac{\gamma^2l}{16G},
-\frac{\tilde{\gamma}^2l}{16G})} \nonumber \\ 
&&~~~~~~ =
\frac{l^2}{4}
\left\{ 
-d(\gamma w)^2-d(\tilde{\gamma}\tilde{w})^2
-(e^{\rho+\ln \frac{4}{\gamma \tilde{\gamma}}}+
e^{-\rho-\ln \frac{4}{\gamma \tilde{\gamma}}})
d(\gamma w)d(\tilde{\gamma}\tilde{w})+(d\rho)^2 
\right \}. 
\end{eqnarray}
In this case $Z_0$ will be taken in $\widetilde{SL}_2({\bf R})$. 
If one considers it in $SL_2({\bf R})$, 
it consists of the elements 
\begin{eqnarray}
e^{-\gamma w J^0}e^{(\rho+\ln \frac{4}{\gamma \tilde{\gamma}})J^1}
e^{-\tilde{\gamma}\tilde{w} J^0}. 
\label{conical Z0}
\end{eqnarray}
The three-geometry which we want to describe is 
$Y_{(-\frac{\gamma^2l}{16G},
-\frac{\tilde{\gamma}^2l}{16G})}$. 
It is obtained from $Z_0$ by the projection $\pi$. 
A comparison of (\ref{conical Z0}) with the Cartan decomposition 
of $SL_2({\bf R})$ given by 
$e^{-(\frac{t}{l}+\theta)J^0}e^{\sigma J^1}
e^{-(\frac{t}{l}-\theta)J^0}$ shows that 
$Z_0$ can be taken in $\widetilde{SL}_2({\bf R})$ 
as the ``cheese cake" with angle $\pi(\gamma+\tilde{\gamma})$. 
See Fig.5.
\begin{figure}[t]
\epsfysize=8cm
\makebox[17cm][c]{\epsfbox{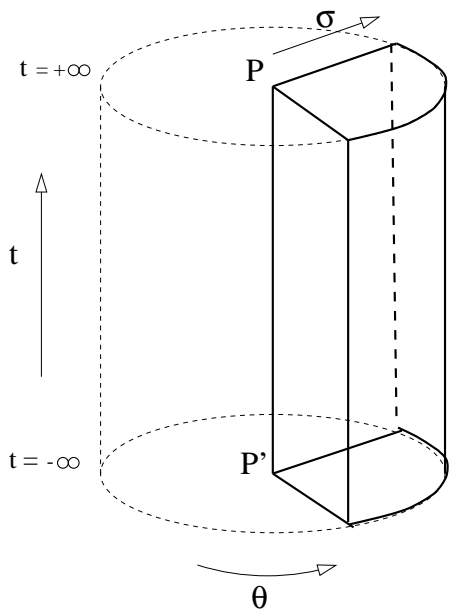}}
\caption{{\small The region $Z_{0}$ in $\widetilde{SL}_2({\bf R})$.
It is the universal covering space of $Y_{(-\frac{\gamma^2l}{16G},
-\frac{\tilde{\gamma}^2l}{16G})}$.}} 
\end{figure}
The ranges are 
$-\infty < t< +\infty, \ 0 \leq \theta <\pi(\gamma+\tilde{\gamma})$ 
and $0 <\sigma < +\infty$. 
The projection $\pi$ causes a conical singularity at $\sigma=0$ 
(Fig.6). 
\begin{figure}[t]
\epsfysize=4.5cm
\makebox[17cm][c]{\epsfbox{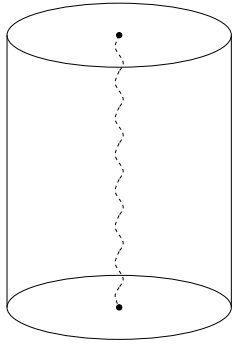}}
\caption{{\small $Y_{(-\frac{\gamma^2l}{16G},
-\frac{\tilde{\gamma}^2l}{16G})}$. The dotted wavy line denotes a
conical singularity which will be added to $Y_{(-\frac{\gamma^2l}{16G},
-\frac{\tilde{\gamma}^2l}{16G})}$ after the completion at $\sigma = 0$.}}
\end{figure}
Mathematically speaking, 
$\pi : Z_0 \rightarrow Y_{(-\frac{\gamma^2l}{16G},
-\frac{\tilde{\gamma}^2l}{16G})}$ is regular but, 
once we make the completion of $Z_0$ at $\sigma=0$ by adding 
the line $PP^{'}$ and extend $\pi$ to this line, 
it becomes singular. This singular mapping causes 
the conical singularity. 
In particular, when $\gamma=\tilde{\gamma}$, 
the appearance of the conical singularity is due to a 
particle with mass $\frac{1-\gamma}{4G}$ sitting at $\sigma=0$ 
\cite{D-J}.

\subsubsection{How can we understand $3d$ black holes 
in two dimensions ?}

    Having obtained the correspondence between the states of 
the three-dimensional quantum gravity and the two-dimensional 
conformal field theory, 
one may ask how the three-dimensional black holes 
can be understood in two dimensions where the 
conformal field lives.

     To investigate this question, it is useful to 
discuss first the case of the conical sigularities. 
The state $|k_{\gamma}\rangle_{in} \otimes 
|\tilde{k}_{\tilde{\gamma}}\rangle_{in}$  
of the conical singularity can be obtained by an operation 
of the corresponding vertex operator on the 
$sl_2(\bf{C})$-invariant vacuum,  
\begin{eqnarray}
|\gamma,\tilde{\gamma} \rangle_{in}= 
\lim_{z,\bar{z} \rightarrow 0} 
e^{ik_{\gamma}X(z)}e^{i\tilde{k}_{\tilde{\gamma}}X(\bar{z})}
|vac \rangle_{in}. 
\label{vertex operator 2}
\end{eqnarray}
Making the holomorphic and anti-holomorphic pieces together, 
we will rewrite the above operator into the form,
\begin{eqnarray}
e^{ik_{\gamma}X(z)}e^{i\tilde{k}_{\tilde{\gamma}}X(\bar{z})}
= 
\displaystyle{
e^{\sqrt{\frac{l}{8G}}(1-\frac{\gamma+\tilde{\gamma}}{2})
X(z,\bar{z})
-\sqrt{\frac{l}{8G}}\frac{\gamma-\tilde{\gamma}}{2}
\widetilde{X}(z,\bar{z})}}, 
\end{eqnarray} 
where $X(z,\bar{z})=X(z)+X(\bar{z})$  
and 
$\widetilde{X}(z,\bar{z})\equiv X(z)-X(\bar{z})$~\footnote{
$\widetilde{X}=\int \ast_2 dX$}. 
If one takes the path-integral formulation of $2d$ CFT, 
the norm of this state, which should be normalized to unity, 
can be written in the following manner : 
\begin{eqnarray}
&&{}_{out}\langle \gamma,\tilde{\gamma} | 
\gamma,\tilde{\gamma} \rangle_{in} 
\nonumber \\ 
&&
= \int DX e^{-S} \cdot 
e^{\sqrt{\frac{l}{8G}}(1+\frac{\gamma+\tilde{\gamma}}{2})
X(\infty)
+\sqrt{\frac{l}{8G}}\frac{\gamma-\tilde{\gamma}}{2}
\widetilde{X}(\infty)}
\cdot 
e^{\sqrt{\frac{l}{8G}}(1-\frac{\gamma+\tilde{\gamma}}{2})
X(0)
-\sqrt{\frac{l}{8G}}\frac{\gamma-\tilde{\gamma}}{2}
\widetilde{X}(0)} 
\label{norm by path-integral 21} 
\end{eqnarray}
where $S$ is the action (\ref{Liouville action}) 
with $\alpha_0=\sqrt{\frac{l}{8G}}$.  
The vertex operators inserted at $0$ and $\infty$ 
respectively create the states of the conical singularity 
from the $sl_2({\bf C})$ invariant in- and out- vacua.   
Choosing the background metric so that   
its curvature concentrates at $\infty$, 
we rewrite the expression 
(\ref{norm by path-integral 21}) in the following form :
\begin{eqnarray}
&&{}_{out}\langle \gamma,\tilde{\gamma} | 
\gamma,\tilde{\gamma} \rangle_{in} 
\nonumber \\ 
&&
=\int DX e^{-\frac{1}{4\pi i}
\int  
\bar{\partial}X \wedge \partial X} \cdot 
e^{-\sqrt{\frac{l}{8G}}(1-\frac{\gamma+\tilde{\gamma}}{2})
X(\infty)
+\sqrt{\frac{l}{8G}}\frac{\gamma-\tilde{\gamma}}{2}
\widetilde{X}(\infty)}
\cdot 
e^{\sqrt{\frac{l}{8G}}(1-\frac{\gamma+\tilde{\gamma}}{2})
X(0)
-\sqrt{\frac{l}{8G}}\frac{\gamma-\tilde{\gamma}}{2}
\widetilde{X}(0)} .  
\nonumber \\ 
~~~~
\label{norm by path-integral 22}
\end{eqnarray}
The Liouville field $X$ treats the conical singularity as an 
insertion of the corresponding vertex operator. This vertex 
operator has the following origin in three-dimensions. 
To explain this, we first remark that 
the ${\bf P}^1$ where the Liouville 
field theory lives can be regarded as that obtained by 
a compactification of the boundary cylinder at infinity.  
This compactification also makes the solid cylinder 
to a three-dimensional ball. 
In such a compactification to a three-ball, 
the conical sigularity located at the center 
must intersect with the ${\bf P}^1$ precisely 
at the two points, $0$ and $\infty$. 
See Fig.7.
\begin{figure}[t]
\epsfysize=5cm
\makebox[17cm][c]{\epsfbox{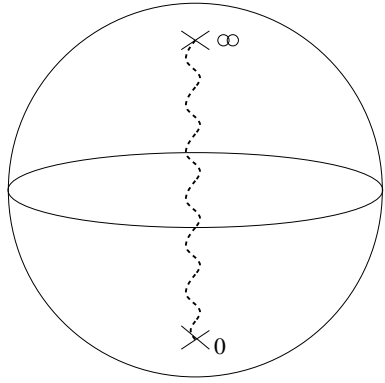}}
\caption{{\small The conical singularity seen by the boundary 
Liouville field $X$}}
\end{figure}
Now, back to the path-integral expression given in 
(\ref{norm by path-integral 22}), we can say that 
these intersections of the conical singularity 
with the boundary sphere are realized as the vertex operators 
in the boundary Liouville field theory. 
In particular, 
the gravitational state of a particle with mass 
$\frac{1-\gamma}{4G}$ sitting at the center of 
$AdS_3$ can be treated by an insertion of 
the corresponding vertex operator.

       Nextly let us discuss about the case of the black holes. 
Basically we follow the same argument as above. 
The black hole state 
(\ref{black hole state in 2d})
can be obtained by an operation of 
the corresponding vertex operator 
on the $sl_2(\bf{C})$-invariant vacuum,  
\begin{eqnarray}
|J,M \rangle_{in}= 
\lim_{z,\bar{z} \rightarrow 0} 
\displaystyle{
e^{(\sqrt{\frac{l}{8G}}+i\frac{r_+}{\sqrt{8Gl}})
X(z,\bar{z})
+i\frac{r_-}{\sqrt{8Gl}}
\widetilde{X}(z,\bar{z})}}
|vac \rangle_{in}. 
\label{vertex operator 1}
\end{eqnarray}
If one takes the path-integral formulation,   
the norm of the black hole state  
can be written in the following manner :   
\begin{eqnarray}
&&{}_{out}\langle J,M | J,M \rangle_{in} 
\nonumber \\ 
&&~~~~~
=\int DX e^{-\frac{1}{4\pi i}
\int  
\bar{\partial}X \wedge \partial X} \cdot 
e^{
-(\sqrt{\frac{l}{8G}}+i\frac{r_+}{\sqrt{8Gl}})X(\infty)
-i\frac{r_-}{\sqrt{8Gl}}
\widetilde{X}(\infty)} \cdot  
e^{(\sqrt{\frac{l}{8G}}+i\frac{r_+}{\sqrt{8Gl}})X(0)
+i\frac{r_-}{\sqrt{8Gl}}\widetilde{X}(0)}.  
\nonumber \\ 
~~~~
\label{norm by path-integral 2}
\end{eqnarray}
This path-integral representation 
also provides some idea 
about our interpretation of the black holes in the 
two dimensions.

                       We first remark that  
the outer horizon of the black hole can not be recognized  
as a two-dimensional object under the Virasoro deformation. 
See Fig.3 for the non-extremal case.
It is treated as a one-dimensional object for the massive black
hole, and as a point for the massless black hole. 
The Virasoro deformation is originally introduced as 
a deformation of the non-compact simply-connected 
region $Z_0$. This region of $SL_2({\bf R})$ 
is the covering space of the exterior of 
the outer horizon of the black hole.  
Under the Virasoro deformation 
what one can recognize as the outer horizon  
is the counterpart of the outer horizon on $SL_2({\bf R})$. 
It is obtained by the completion of $Z_0$ at $\sigma=0$.  
We can easily see that these completions  
for the cases of the massive extremal 
and massless black holes are respectively 
given by adding one-dimensional and zero-dimensional 
objects. Being zero-dimensional in the massless case 
may be understood as a degeneration from the massive case.

                         We hope to explain the geometrical 
origin of the insertion of the vertex operator in 
(\ref{norm by path-integral 2}). 
Let us examine first the compactification of the boundary cylinder 
of the black hole $X^{+}_{(J,M)}$ to the ${\bf P}_1$.   
$t=\pm \infty$ circles of the boundary cylinder are mapped to 
the two points $0$ and $\infty$ of the ${\bf P}_1$. 
This compactification of the boundary 
also provide a compactification of the bulk geometry. 
Now we will ask what it gives rise to. 
We can assume reasonably that $t=\pm \infty$ slices 
of $X_{(J,M)}^+$ are compactified to $0$ and $\infty$ of the 
${\bf P}_1$. If one accepts this assumption, 
the outer horizon must be compactified to $0$ and $\infty$ 
of ${\bf P}_1$. 
This is because the $t=\pm \infty$ slices of $X_{(J,M)}^+$ 
include the outer horizon after the completion. See Fig.3.   
This shows that the intersections of the black hole with the 
boundary sphere are realized by the vertex operators as described 
in (\ref{norm by path-integral 2}). See Fig.8.

\begin{figure}[t]
\epsfysize=5cm
\makebox[17cm][c]{\epsfbox{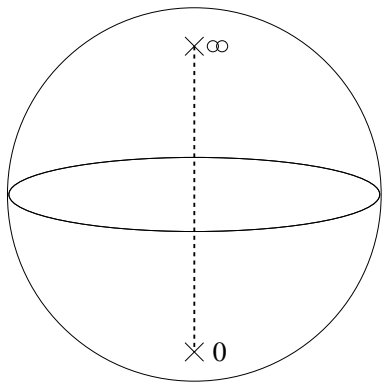}}
\caption{{\small The black hole seen by the boundary Liouville field $X$}}
\end{figure}

\section{Towards Conformal Field Theory On The Horizon}

                  To provide a microscopic description of black holes 
in quantum gravity or string theory is a very impressive and enlightening 
issue. For the three-dimensional black holes such a microscopic description 
has been proposed 
by Carlip \cite{Carlip1} 
and Strominger 
\cite{Strominger1,Maldacena-Strominger}. 
It states that microscopic states of the black holes 
are the states of a conformal field theory on the horizon. 
Although at present we do not know 
exactly what conformal fields do live on the horizon,  
their proposal makes it possible to give  
statistical mechanical explanations of 
the thermodynamical properties of the black hole.

               Receiving their proposal, 
many questions will arise. 
First of all, can one identify the boundary Liouville field 
theory with a conformal field theory on the horizon ? 
Our answer is NO.  
As explained at the end of the last section, 
the Virasoro deformation of the exterior of the outer horizon   
can not recognize the horizon as a two-dimensional object.  
It can recognize the horizon only as a one-dimensional object. 
This means that 
the Virasoro algebra obtained from the deformation 
cannot be the Virasoro algebra on the horizon.   
On the other hand, 
this deformation provides the asymptotic Virasoro 
algebra, which generators are gravitons in bulk and 
are identified with the excitations of the boundary Liouville field. 
This field theory has the continuous spectrums. 
Therefore these two conformal field theories are by no means 
equivalent to each other.

           If one accepts this answer, 
one might be led to another question : 
What conformal field theory is living on the horizon ? 
To this question we can not answer definitely.  
Here we would like to propose two possible descriptions of this 
conformal field theory. 
The first one is based on a simple hypothesis : 
The Virasoro algebra of the horizon conformal field theory 
should generate the Virasoro deformation of the horizon.   
To work this hypothesis, 
we need, first of all, to treat the outer horizon 
as a {\it two-dimensional} object. 
If one takes the Chern-Simons gravity viewpoint, 
it is doubtful whether one can handle the outer horizon 
as a two-dimensional object. To be successful,  
presumably one needs the string theory viewpoint.  
Namely,   
if the two-dimensional horizon is identified with 
a macroscopic string in $SL_2({\bf R})$, 
the horizon conformal field theory would be string theory 
on this background.
It describes quantum fluctuations of the macroscopic string 
and will lead the microscopic description of the black holes.  
But it is not clear, 
at least for us, 
to what extent this string theory is analyzable.

              The second possible description 
of the horizon conformal field theory is based on 
the hypothesis : The Virasoro deformation of the region 
between the inner and outer horizons should lead 
the Virasoro algebra on the horizon conformal field theory.    
This hypothesis can be regarded as an analogue which we 
used in the construction of the boundary Liouville field 
theory. In particular it may allow us to follow the similar 
step as we took for the study of the exterior of the 
outer horizon.

            To be explicit, 
let us consider the region between the inner and outer horizons 
of the non-extremal BTZ black hole $X_{(J,M)}$. We will call it 
$X_{(J,M)}^-$  
\footnote{Explicitly,  
$X_{(J,M)}^-$ $\equiv$ 
$\left\{ \left.(t,\phi,r) 
\in X_{(J,M)} \right| r_-<r<r_+ \right\}$.}.   
A covering space of $X_{(J,M)}^-$ is given  
by a non-compact simply-connected region of $SL_2({\bf R})$ 
consisting of the group elements  
\begin{equation}
e^{-\varphi J^2} e^{\sigma J^0} e^{\psi J^2}, 
\label{Z1 non-ex}
\end{equation}
where the ranges of $\varphi,\psi$ and $\sigma$ are taken 
$-\infty< \varphi, \psi <+\infty$ and $0<\sigma<\pi$. 
Let us call this region $Z_1$. 
The projection can be described by 
\begin{eqnarray}
\frac{r_+ +r_-}l \left(\frac{t}{l}+\phi\right) 
= \varphi,~~~
\frac{r_+ -r_-}l \left(\frac{t}{l}-\phi\right) 
= \psi,~~~  
\frac{2r^2-r_+^2 -r_-^2}{r_+^2 -r_-^2} 
= \cos\sigma, 
\end{eqnarray}
together with the identification $\phi \sim \phi+2\pi$. 
The induced metric of $Z_1$ has the form 
\begin{equation}
{ds}^2_1 = \frac{l^2}4 \left\{(d\varphi)^2+(d\psi)^2
          -2\cos\sigma~d\varphi d\psi -(d\sigma)^2\right\}, 
\end{equation}
and is isometric to $ds_{X_{(J,M)}^-}^2$. 
Therefore $Z_1$ is indeed the covering space.  
$X_{(J,M)}^-$ is identified with the quotient 
of $Z_1$, 
\begin{equation}
X^-_{(J, M)} = Z_1/\sim, 
\end{equation}
where the equivalence relation is 
$(\varphi, \psi, \sigma) \sim 
(\varphi+2\pi\frac{r_+ +r_-}l , \psi-2\pi\frac{r_+ -r_-}l, \sigma)$.
As is done in the study of $Z_0$, 
it is convenient to rewrite 
the group element (\ref{Z1 non-ex}) in the form 
\begin{equation}
e^{-\sqrt{\frac{b_0}{c/24}}w J^2} 
 e^{(\rho-\frac12\ln\frac{b_0 \tilde{b}_0}{(c/6)^2}) J^0} 
e^{\sqrt{\frac{\tilde{b}_0}{c/24}}\tilde{w} J^2}~~~
(\equiv h_1(w,\tilde{w},\rho)). 
\label{h1 non-ex} 
\end{equation}
Description of $Z_1$ in terms of the Chern-Simons gravity 
is given by the flat connection 
$A_1=(A_1^{(+)},A_1^{(-)})$ which is determined 
by the prescription (\ref{def of A0}) using    
a decomposition of $h_1$  
into the following $SL_2({\bf R})_{\pm}$-valued functions 
$h^{(\pm)}_1$ : 
\begin{eqnarray}
{h}_1^{(+)} 
= e^{-\frac12(\rho-\ln\frac{b_0}{c/6}) J^0}
             e^{\sqrt{\frac{b_0}{c/24}}w J^2},~~~~
{h}_1^{(-)} 
= e^{\frac12(\rho-\ln\frac{\tilde{b}_0}{c/6}) J^0}
          e^{\sqrt{\frac{\tilde{b}_0}{c/24}}\tilde{w} J^2}. 
\label{h1pm non-ex}
\end{eqnarray}
A slight comparison between $h_1^{(\pm)}$ and  
$h^{(\pm)}$ given in (\ref{hpm non-ex}) shows that 
the flat connection $A_1$ is gauge-equivalent to 
$A_0$ (\ref{A0 non-ex}), 
the flat connection which 
describes $Z_0$, the covering space of 
the exterior of the horizon :   
\begin{eqnarray} 
A_1=A_0^g, 
\end{eqnarray}   
where the gauge transformation 
$g=(g^{(+)},g^{(-)})$ is given by 
$g^{(+)} 
= e^{-\frac12(\rho-\ln\frac{b_0}{c/6}) J^0}
             e^{\frac12(\rho-\ln\frac{b_0}{c/6}) J^1}$ 
and 
$g^{(-)} 
= e^{\frac12(\rho-\ln\frac{\tilde{b}_0}{c/6}) J^0}
             e^{-\frac12(\rho-\ln\frac{\tilde{b}_0}{c/6}) J^1}$. 
This gauge equivalence should be understood as a formal one 
because the allowed ranges of the coordinates are different 
from each other. 
(For $A_0$ (\ref{A0 non-ex}) the range of $\rho$ is  
$\frac12\ln\frac{b_0 \tilde{b}_0}{(c/6)^2}< \rho <+\infty$ 
while for $A_1$, 
$\frac12\ln\frac{b_0 \tilde{b}_0}{(c/6)^2}< \rho 
<\pi + \frac12\ln\frac{b_0 \tilde{b}_0}{(c/6)^2}$ .) 
Nevertheless, if one takes the existence of such a 
gauge transformation seriously, 
one may arrive at an idea that a deformation of $Z_1$ 
could be constructed in such a manner 
that it is related with the Virasoro deformation of $Z_0$ 
by an appropriate gauge transformation.   
Thus obtained deformation will define 
the Virasoro deformation of $Z_1$ and thereby 
that of $X_{(J,M)}^-$, the region between the inner 
and outer horizons of the black hole. The corresponding 
conformal field theory will be regarded 
as the horizon conformal field theory.

                 The second description  
of the horizon conformal field theory may provide 
another interpretation of the surprising conjecture 
made by Maldacena \cite{Maldacena}, 
which may be understood in our situation 
as a correspondence between the boundary and horizon 
conformal field theories under a suitable 
renormalization group flow.  The second description, 
if it is correct, implies a possibility of a realization 
of this correspondence in terms of gauge transformation.

~

~

We thank S. Mano for useful discussions.

~

\section*{Note Added}

After this work was completed we received a preprint 
\cite{N-N} in which the content of 3.3 is also discussed.


\end{document}